\documentclass[12pt]{article}

\usepackage{amsmath}
\usepackage[round]{natbib}
\usepackage{amssymb,color}
\usepackage{appendix}
\usepackage{amsthm}
\usepackage{enumerate}
\usepackage{mathtools} 
\usepackage{hyperref}
\usepackage{subcaption}

\usepackage[linewidth=1pt]{mdframed}
\usepackage{lipsum}
\usepackage{listings}
\mathtoolsset{showonlyrefs=true}  

\DeclareMathOperator*{\argmin}{arg\,min}
\DeclareMathOperator*{\arginf}{arg\,inf}
\usepackage{caption}

\usepackage[margin=0.7in]{geometry}

\numberwithin{equation}{section}
\usepackage{comment}

\usepackage{titling}
\newtheorem{thm}{Theorem}[section]

\newtheorem{defi}{Definition}[section]

\begin{document}

\author{Daeyung Gim \thanks{gbe375@snu.ac.kr} and  Hyungbin Park \thanks{hyungbin@snu.ac.kr, hyungbin2015@gmail.com}  \\ \\ \normalsize{Department of Mathematical Sciences and Research Institute of Mathematics} \\ 
	\normalsize{Seoul National University}\\
	\normalsize{1, Gwanak-ro, Gwanak-gu, Seoul, Republic of Korea} 
}
\title{A deep learning algorithm for optimal investment strategies}

\maketitle

\abstract{
This paper treats the Merton problem  how to invest in safe assets and risky assets to maximize an investor's utility, given by investment opportunities modeled by a $d$-dimensional state process. The problem is represented by a partial differential equation with optimizing term: the Hamilton--Jacobi--Bellman equation. The main purpose of this paper is to solve partial differential equations derived from the Hamilton--Jacobi--Bellman equations with a deep learning algorithm: the Deep Galerkin method, first suggested by \cite{sirignano2018dgm}. We then apply the algorithm to get the solution of the PDE based on some model settings and compare with the one from the finite difference method.
}

\section{Introduction} \label{sec1}

Consider the following expected utility maximization problem:
\begin{equation}
	\max_{(\pi_u)_{u\geq t}} \frac{1}{p} \, \mathbb{E}\left[(X^{\pi}_T)^p \, | \, X_t=x, \, Y_t=y\right],
\end{equation}
where $\pi$ is a portfolio, $X^{\pi}$ a wealth process and $Y$ a state variable with the utility function $(1/p)x^p=:U(x)$. This kind of problem is first suggested by \cite{merton1969lifetime}, which is the most fundamental and pioneering in economics. The Merton problem has played as a key for an investor's wealth allocation in several assets under some market circumstances. Since then there have been lots of studies about Merton problem under various conditions. \cite{benth2003merton} studied Merton problem under the Black-Scholes setting by using the OU type stochastic volatility model. \cite{kuhn2010optimal} studied optimizing portfolio of Merton problem under a limit-ordered market in view of a shadow price. The research on the optimal investment based on inside information and drift parameter uncertainty was conducted by \cite{danilova2010optimal}. \cite{nutz2010opportunity} studied the utility maximization in a semimartingale market setting with the opportunity process. \cite{hansen2013optimal} suggested an optimal investment strategies with investors' partial and private information. \cite{pedersen2017optimal} applied the Lagrange multiplier to solve nonlinear mean-variance optimal portfolio selection problem. Also there was research on the optimal portfolio strategies using over-reaction and under-reaction by \cite{callegaro2017optimal}. \cite{liang2020robust} researched a robust Merton problem using the constant relative/absolute risk aversion utility functions under the time-dependent sets of confidence.

In this paper we follow the overall market setting in \cite{guasoni2015static} and induce the so-called Hamilton--Jacobi--Bellman equation under time variable $t$, variable $x$ representing wealth process and variable $y=(y_1,\ldots,y_d)$ from the $d$-dimensional state variable. We can optimize the portfolio by means of finding a solution to the HJB equation. By using some properties including homotheticity and concaveness, we eliminate the optimizing term to change the HJB equation into a nonlinear partial differential equation.

Under this circumstance we face with the problem of solving nonlinear PDEs. Because in general most PDEs do not have analytic solutions, there exists several well-known numerical tools. These classical approaches can be found in \cite{achdou2005computational} and \cite{burden2010numerical}.

At the same time there has been some studies about solving PDEs with a deep neural network. \cite{lee1990neural}, \cite{lagaris2000neural} suggested the neural network algorithm on a fixed mesh. \cite{malek2006numerical} also suggested the numerical hybrid DNN optimizing method. However in case of the higher dimension of PDEs, these grid-based methods would be computationally inefficient: a curse of dimensionality.

Recently there have been several researches to get rid of the curse of dimensionality using machine learning techniques. \cite{han2018solving} and \cite{weinan2019multilevel} suggested a deep backward stochastic differential equation method with the Feynman--Ka\v{c} formula.

The deep learning algorithm mainly used in this paper is the Deep Galerkin method suggested by \cite{sirignano2018dgm}. It is computationally efficient since there does not need to make any mesh or grid. We define a loss functional to minimize $L^2$-norm about the desired differential operator and other conditions from the PDE. To make the loss small enough as we want, we sample random points from the domain and optimize by means of stochastic gradient descent. After deriving surfaces, we also apply the finite difference method(FDM) in order to compare surfaces from both algorithms: DGM and FDM. For further research on the Deep Galerkin method, see \cite{al2018solving} and \cite{al2019applications}.

This paper is organized as follows. In \autoref{sec2}, we start by describing the general setting of this paper, and induce the partial differential equation with optimizing term: the HJB equation. The Deep Galerkin method algorithm and neural network approximation theorem from \cite{sirignano2018dgm} are presented in \autoref{sec3}, with some part of code for each step of DGM algorithm. Numerical test of the algorithm is presented in \autoref{sec4}. Specifically, we model $2$ dimensional state process by the OU process and the CIR process, return process by the Heston model. Then we use the calibrated parameters from \cite{crisostomo2014analyisis} and \cite{mehrdoust2020calibration}. We display the solution surface at each fixed time in some pre-determined domain of the state variable. We finally analyze surfaces from the Deep Galerkin method and those from the finite difference method. Conclusions can be found in \autoref{sec5}, and proofs of neural network approximation theorem are in appendix A.

\section{Optimal Investment Problem}  \label{sec2}

In the case that an economic agent is in time interval $[0,T]$, the problem is that he or she has to decide how to invest in several risky assets or safe assets as time goes by, starting with the initial wealth. This problem was first suggested by Merton in the 1960s: Merton problem, known as a utility maximization problem. The aim of the agent is to establish a portfolio strategy in such a way of maximizing utility under some conditions. In this section we describe the general setting of this paper, and induce the HJB equation. We finally reach to a nonlinear PDE by using some properties. The above problem is equivalent to a matter of finding a solution of the equation.

\subsection{Market with the Merton Problem}

We first start by describing market with the following framework. Assume that the market has $n+1$ assets $S^{(0)},S^{(1)},\ldots,S^{(n)}$, where $S^{(0)}$ is safe and $S^{(1)},\ldots,S^{(n)}$ are risky. One can make a decision to the investment by a $d$-dimensional state variable $Y=(Y^{(1)},\ldots, Y^{(d)})$ satisfying:
\begin{equation} \label{Y_SDE} 
	dY_t=b(Y_t)\,dt+a(Y_t)\,dW_t,
\end{equation}
where $W=(W^{(1)},\ldots,W^{(d)})$ denotes a standard Brownian motion.

Let $r$ be the interest rate, $\mu$ be the excess returns, and $\sigma$ be the volatility matrix. We also assume that the prices of the assets satisfy:
\begin{equation}   \label{eqn:S0}       
	dS^{(0)}_t = r S^{(0)}_t dt,
\end{equation}
\begin{equation}   \label{eqn:S}
    \dfrac{dS^{(i)}_t}{S^{(i)}_t} = r dt + dR^{(i)}_t \qquad 1 \leq i \leq n,
\end{equation}
where $R=(R^{(1)},\ldots, R^{(n)})$ denotes the cumulative excess return satisfying:
\begin{equation} \label{eqn:R}
    dR^{(i)}_t = \mu_i(Y_t) dt + \sum_{j=1}^{n} \sigma_{ij}(Y_t) \, dZ^{(j)}_t \qquad 1 \leq i \leq n.
\end{equation}
$\rho=(\rho_{ij})=d\left\langle Z, \, W \right\rangle_t / dt$ denotes the cross correlations between the $n$-dimensional Brownian motion $Z$ and $W$. $\Sigma = \sigma\sigma^T = d\left\langle R, \, R \right\rangle_t / dt$ is the matrix of quadratic covariance of returns, and $ \Upsilon = \sigma\rho a^T = d\left\langle R, \, Y \right\rangle_t / dt$ denotes the correlation between the return and the state process.

In the market, an investor buys the risky assets by a portfolio $\pi=(\pi^{(1)}_t,\ldots,\pi^{(n)}_t)_{t\geq0}$. The wealth process $X^{\pi}=(X^{\pi}_t)_{t\geq0}$ corresponding to the portfolio satisfies
\begin{equation} \label{X_SDE}
    \dfrac{dX^{\pi}_t}{X^{\pi}_t} = r \, dt + \pi^T_t \, dR_t, \quad X^{\pi}_0 \geq 0.
\end{equation}
Observe first that the portfolio process $(\pi_t)_{t\geq0}$ is $\mathcal{F}_t$-measurable, where the filtration $\mathcal{F}=(\mathcal{F}_t)_{t\geq0}$ is generated by the return $R$ and state variable $Y$. It might be clear in light of the investor's eyes: he or she has all informations about state and asset return from time $t=0$ to the current time. Note also the portfolio process is integrable with respect to the return process $R$.
By the Merton problem, we assume the investors' utility function is defined by the following:
\begin{equation}
	U(x) = \dfrac{1}{p}x^{p}, \quad 0<p<1.
\end{equation}
For fixed wealth $x$ and state $y=(y_1,\ldots,y_d)$ satisfying \eqref{Y_SDE} and \eqref{X_SDE}, our aim is to maximize the conditional expectation of terminal wealth utility given wealth and state at time $t$, that is
\begin{equation}
    \max_{(\pi_u)_{u\geq t}} \frac{1}{p} \, \mathbb{E}\left[(X^{\pi}_T)^p \, | \, X_t=x, \, Y_t=y\right].
\end{equation}
\subsection{The Hamilton--Jacobi--Bellman Equation}

Now we substitute the problem of utility maximization to that of solving the PDE, namely the Hamilton--Jacobi--Bellman equation. There needs to be some definitions before approaching to the HJB equation.
\begin{defi}
	A portfolio process $\pi=(\pi_t)_{t\geq0}$ is called an \textbf{admissible} portfolio if
	\begin{itemize}
		\item For every $t\in[0,T]$ and $(x,y) \in D \subset \mathbb{R} \times \mathbb{R}^{d}$, $\pi(t,x,y) \in U$, where $U\subset\mathbb{R}$ is a fixed subset.
		\item For any given initial points $(t,x)$ and $y=(y_1,\ldots,y_d)$, the following SDE has a unique solution:
		\begin{equation}    \label{X_SDE_s}       
			\begin{aligned}
				dX^{\pi}_s &= r X^{\pi}_s \, ds + \pi^T_s \, dR_s, \\
				X^{\pi}_t &=x.
			\end{aligned}
		\end{equation}
		\item For any given initial point $(t,y)=(t,y_1,\ldots,y_d)$, the following SDE has a unique solution:
		\begin{equation}    \label{Y_SDE_s}       
			\begin{aligned}
				dY_s &= b(Y_s) \, ds + a(Y_s) \, dW_s, \\
				Y_t &=y.
			\end{aligned}
		\end{equation}
	\end{itemize}
\end{defi}

By now we assume the portfolio $\pi$ is admissible.
\begin{defi} Let $U$ be an investor's utility function.
    \begin{itemize}
    	\item For each $\pi$, we define the \textbf{expected value function} $\mathcal{V}^{\pi}$ as
    	\begin{equation}
    		\mathcal{V}^{\pi}(t,x,y)=\mathbb{E}[U(X^{\pi}_T) | X_t=x, \, Y_t=y],
    	\end{equation}
        given \eqref{X_SDE_s} and \eqref{Y_SDE_s}.
        \item We define the \textbf{optimal value function} $V$ as
        \begin{equation}
        	V(t,x,y)=\sup_{\pi}\mathcal{V}^{\pi}(t,x,y).
        \end{equation}
    \end{itemize}
\end{defi}

The following theorem justifies a conversion from the way of finding optimal portfolio to that of solving PDEs having optimizing term. Heuristic process for deriving the HJB equation is in chapter 19, \cite{bjork2009arbitrage}, in the way of limiting procedures in dynamic programming.

\begin{thm} \label{thm:HJB}
	Assume the following.
	\begin{itemize}
		\item The market has a safe asset $S^{(0)}$ whose dynamics is expressed in \eqref{eqn:S0}.
		\item The market has $n$ risky assets satisfying \eqref{eqn:S}, with the return process $R$ following the diffusion \eqref{eqn:R}.
		\item There exists an optimal portfolio $\hat{\pi}=(\hat{\pi}^{(1)}_t,\ldots,\hat{\pi}^{(n)}_t)_{t\geq0}$.
		\item The optimal value function $V$ is regular, that is, $V \in C^{1,2,2}$ with respect to $(t,x,y)$, $y=(y_1,\ldots,y_d)$.
	\end{itemize}
	Then the following hold:
	\begin{enumerate}
		\item $V$ satisfies the Hamilton--Jacobi--Bellman equation
		\begin{equation}
			\begin{aligned}
				V_t + b^T (\nabla_y V) + \dfrac{1}{2} \, \text{tr} [a^T (\nabla^2_y V)\,a] + rxV_x \\+ \sup_{\pi} \left[ \pi^T (\mu V_x+\Upsilon(\nabla_y V_x))x + \dfrac{1}{2}x^2 V_{xx} \pi^T \Sigma \pi \right]&=0, &(t,x,y) \in [0,T]\times D, \\
				V(0,x,y)&=U(x), &(x,y) \in D.  \\
			\end{aligned}
		\end{equation}
		\item An optimizing term in the above equation can be achieved by $\pi=\hat{\pi}$:
		\begin{equation}
			\sup_{\pi}\left[ \pi^T (\mu V_x+\Upsilon(\nabla_y V_x))x + \dfrac{1}{2}x^2 V_{xx} \pi^T \Sigma \pi \right]=\hat{\pi}^T (\mu V_x+\Upsilon(\nabla_y V_x))x + \dfrac{1}{2}x^2 V_{xx} \hat{\pi}^T \Sigma \hat{\pi}.
		\end{equation}
	\end{enumerate}
	
\end{thm}

If we define the optimal value function as
\begin{equation}   \label{eqn:valuefunc}       
	V(t,x,y_1,\ldots,y_d) = \sup_{(\pi_u)_{u\geq t}}  \, \mathbb{E}\left[\frac{1}{p}(X^{\pi}_T)^p \, | \, X^{\pi}_t=x, Y_t^{(1)}=y_1,\ldots,Y_t^{(d)}=y_d\right],
\end{equation}
by Theorem \ref{thm:HJB} with the It\^{o} formula, one can derive the Hamilton--Jacobi--Bellman equation from \eqref{eqn:valuefunc}:\\
\begin{equation}   \label{eqn:HJB_V}       
\begin{split}
	V_t + b^T (\nabla_y V) + \dfrac{1}{2} &\, \text{tr} [a^T (\nabla^2_y V)\,a] + rxV_x \\
	 +& \sup_{\pi} \left[ \pi^T (\mu V_x+\Upsilon(\nabla_y V_x))x + \dfrac{1}{2}x^2 V_{xx} \pi^T \Sigma \pi \right]=0,
\end{split}
\end{equation}
where the terminal condition of \eqref{eqn:HJB_V} is $V(T,\,x,\,y)=(1/p)\,x^p$. $\nabla_y V=(V_{y_1},\ldots,V_{y_d})$ and $\nabla^2_y V=\left(V_{y_i y_j}\right)_{1\leq i,j \leq d}$ stand for the gradient and the Hessian of $V$ with respect to $y=(y_1,\ldots,y_d)$, respectively. Because of the concaveness of $V$ in $x$ and $\sup_{\pi}(\pi^T b+\frac{1}{2}\pi^T A \pi)=-\frac{1}{2}b^T A^{-1}b$ for negative definite matrix $A$, \eqref{eqn:HJB_V} becomes
\begin{equation}   \label{eqn:PDE_V}       
\begin{split}
    V_t + b^T (\nabla_y V) + \dfrac{1}{2} &\, \text{tr} [a^T (\nabla^2_y V)\,a] + rxV_x \\
     -& (\mu V_x+\Upsilon(\nabla_y V_x))^T \dfrac{\Sigma^{-1}}{2V_{xx}}(\mu V_x+\Upsilon(\nabla_y V_x))=0,
\end{split}
\end{equation}
with the corresponding optimal portfolio is
\begin{equation}   \label{eqn:port_V}       
	\pi=\pi(t,x,y_1,\ldots,y_d)=-\dfrac{1}{xV_{xx}}\Sigma^{-1}(\mu V_x+\Upsilon(\nabla_y V_x)).
\end{equation}
Since the utility function is homothetic, we define the reduced value function $u$ as
\begin{equation} \label{eqn:V_to_u}
	V(t,x,y_1,\ldots,y_d)=\dfrac{1}{p} \,x^p \, u(t,y_1,\ldots,y_d).
\end{equation}
If we put \eqref{eqn:V_to_u} into \eqref{eqn:PDE_V} and divide each component by $x^p$, \eqref{eqn:PDE_V} becomes
\begin{equation}   \label{eqn:PDE_u}       
\begin{split}
	u_t + (b^T -& q \mu^T \Sigma^{-1} \Upsilon)\nabla_y u + \dfrac{1}{2} \, \text{tr} [a^T (\nabla^2_y u)\,a] \\ + &(pr-\dfrac{q}{2}\mu^T \Sigma^{-1} \mu)u - \dfrac{q}{2u}(\nabla_y u)^T\Upsilon^T\Sigma^{-1}\Upsilon(\nabla_y u)=0,
\end{split}
\end{equation}
where the terminal condition of \eqref{eqn:PDE_u} is $u(T,y_1,\ldots,y_d)=1$. In \eqref{eqn:PDE_u}, we set $q=p/(p-1)$ for simplicity. Also the following is the reduced optimal portfolio:
\begin{equation}   \label{eqn:port_u}       
	\pi(t,y_1,\ldots,y_d) = \dfrac{1}{1-p} \left( \Sigma^{-1}\mu + \Sigma^{-1}\Upsilon(\nabla_y u)\dfrac{1}{u} \right).
\end{equation}

\section{Deep Galerkin Method}  \label{sec3}

Now we investigate how to solve the PDEs such as \eqref{eqn:PDE_u}. Since only few PDEs have analytic solutions, there are well-known numerical tools including the Monte Carlo method exemplified by the Feynman--Ka\v{c} theorem and the finite difference method. However one of the most difficult facts is a curse of dimensionality. In particular in grid-based numerical methods, the number of mesh points grows explosively as the dimension goes higher, so \cite{sirignano2018dgm} suggest a DNN-based algorithm for approximating solution of PDEs: the Deep Galerkin method(DGM), such that there is no need to make any mesh.

With the parametrized deep neural network, say $f$, a loss functional $f \mapsto J(f)$ is defined to minimize $L^2$-norm about the desired differential operator and terminal condition. To make the loss small enough as we want, the network samples random points from the pre-determined domain and is optimized by means of the stochastic gradient descent. In this section we first introduce the DGM algorithm. We then state the approximation theorem in order to justify this new algorithm.

\subsection{Algorithm}

Let $u=u(t,y)$ be an unknown function which satisfies the PDE:

\begin{equation}   
    \begin{aligned}
    \partial_t u(t,y)+\mathcal{L}u(t,y)&=0, &(t,y) \in [0,T]\times D, \\
    u(T,y)&=u_T(y), &y \in D, \label{eqn:pde_form} \\
    \end{aligned}
\end{equation}
where $D \subset \mathbb{R}^d$. Our aim is to express the solution of \eqref{eqn:pde_form} as a neural network function $f=f(t,y;\theta)$ in place of $u$. $\theta=(\theta^{(1)},\cdots,\theta^{(K)})$ denotes a vector of network parameters.

Define a loss functional $J:=J_1+J_2$ with
\begin{equation}  
    \begin{aligned}
        J_1(f)&:=\lVert \partial_t f(t,y;\theta) +\mathcal{L}f(t,y;\theta) \rVert ^2 _{[0,T]\times D,\nu_1} \\
        J_2(f)&:=\lVert f(T,y;\theta)-u_T (y) \rVert ^2 _{D,\nu_2} \label{losses} \\
    \end{aligned}
\end{equation}
Note that all above terms are expressed in terms of $L^2$-norm, that is, $\lVert h(y) \rVert^2_\mathcal{Y,\nu}=\int_\mathcal{Y}\,\lvert h(y) \rvert^2 \nu(y) dy$. Each functionals $J_1$ and $J_2$ determine that how well the approximation has conducted in view of the PDE differential operator and terminal condition. The aim is to find a parameter $\hat{\theta}$ in such a way of minimizing $J(f)$, equivalently,
\begin{equation}
	\hat{\theta}=\argmin_{\theta} J(f(t,y;\theta)).
\end{equation}
As the error $J(f)$ goes smaller, the approximated function $f$ would get closer to the solution $u$. Hence $f(t,y;\hat{\theta})$ might be the best approximation of $u(t,y)$.

\smallskip
The algorithm of DGM is as follows:
\begin{enumerate}
	\item Set initial values of $\theta_0=(\theta^{(1)}_0,\cdots,\theta^{(K)}_0)$ and determine the learning rate $\beta_n$.
	\item Sample random points $(t_n,y_n)$ in $[0,T]\times D$ according to probability density $\nu_1$. Likewise, pick random points $w_n$ from $D$ with density $\nu_2$.
	\item Calculate the $L^2$-error for the randomly sampled points $s_n=\{ (t_n,y_n),w_n\}$:
	\begin{equation}
		L(\theta_n,s_n)=((\partial_t +\mathcal{L})f(t_n,y_n;\theta_n) )^2+(f(T,w_n;\theta_n)-u_T(w_n))^2.
	\end{equation}
	\item Use the stochastic gradient descent at $s_n$:
	\begin{equation}
		\theta_{n+1}=\theta_n-\beta_n\nabla_\theta L(\theta_n,s_n).
	\end{equation}
	\item Repeat until $\lVert \theta_{n+1}-\theta_n\rVert$ is small enough.
\end{enumerate}
\medskip
The following is some part of code for each step of DGM algorithm:

\begin{mdframed}[leftmargin=3pt,rightmargin=3pt]
	\begin{lstlisting}
 # 1-1. Initializing the neural network parameter
 oper_init = tf.global_variables_initializer()

 # 1-2. Initializing the learning rate
 lrn_rate = tf.train.exponential_decay(init_lrn_rate, glob_step, dec_step, dec_rate, staircase=True)
 optimizer = tf.train.AdamOptimizer(lrn_rate).minimize(loss_tnsr)
 
 # 2-1. Generating random samples : interior of the domain
  t_int = np.random.uniform(low=0, high=T, size=[nSim_int,1])
 y1_int = np.random.uniform(low=y1_low, high=y1_high, size=[nSim_int,1])
 y2_int = np.random.uniform(low=y2_low, high=y2_high, size=[nSim_int,1])
 
 # 2-2. Generating random samples : terminal condition
  t_ter = T * np.ones(nSim_ter,1)
 y1_ter = np.random.uniform(low=y1_low, high=y1_high, size=[nSim_ter,1])
 y2_ter = np.random.uniform(low=y2_low, high=y2_high, size=[nSim_ter,1]) 
 
 # 3. Calculating L^2-error of differential operator / terminal condition
  # differential operator
 J1 = tf.reduce_mean(tf.square(diff_u))
  # terminal condition
 J2 = tf.reduce_mean(tf.square(fitted_ter - target_ter))
  J = J1 + J2 
 
 # 4. Stochastic gradient descent step
 for k in range(steps_per_sample):
     loss, J1, J2, k = sess.run([loss_tnsr, J1_tnsr, J2_tnsr, optimizer], feed_dict={t_int_tnsr:t_int, y1_int_tnsr:y1_int, y2_int_tnsr:y2_int, t_ter_tnsr:t_ter, y1_ter_tnsr:y1_ter, y2_ter_tnsr:y2_ter})
	\end{lstlisting}
\end{mdframed}

\subsection{Neural Network Approximation}

The following neural network approximation theorem is stated in \cite{sirignano2018dgm}. In other words, there exists a collection of approximated neural network functions that converges to a solution of quasilinear parabolic PDEs.
\begin{thm} \label{thm:DGM}
	Define $\mathfrak{C}^n$ as a collection of DNN functions with $n$ hidden neurons in a single hidden layer. Assume $u=u(t,y)$ be an unknown solution for \eqref{eqn:pde_form}. Under certain conditions in \cite{sirignano2018dgm}, there exists a neural network function $f^n$ with $n$ hidden neurons such that the following hold:
	\begin{enumerate}
		\item $J(f^n) \to 0$ as $n \to \infty$,
		\item $f^n \xrightarrow{strongly} u$ in $L^{\rho}([0,T]\times D)$ as $n \to \infty$, where $\rho<2$.
	\end{enumerate}
\end{thm}
Some part of proofs for our formulation in this paper is in appendix A. Further details including conditions and proofs are in section 7 and appendix A in \cite{sirignano2018dgm}.

\section{Numerical Test}  \label{sec4}

The key purpose of this section is to solve \eqref{eqn:PDE_u} with the Deep Galerkin method and compare the numerical solution with the one derived by the well-known finite difference method.

\subsection{Model Settings}

We first set some specific settings of the market model. For our experiment we assume that there are two ways of decision for trading, i.e., 2 dimensional state variable $Y=(Y^{(1)},Y^{(2)})$. Let $Y^{(1)}$ be the Ornstein-Uhlenbeck(OU) process and $Y^{(2)}$ be the Cox-Ingersoll-Ross(CIR) process. This state variable is expressed by the following matrix form:
\begin{equation}
	\left(\begin{array}{c}
		dY_{t}^{(1)} \\ dY_{t}^{(2)}
	\end{array}\right)
	=
	\left(\begin{array}{c}
		\theta^{(1)}(k^{(1)}-Y^{(1)}_t) \\ \theta^{(2)}(k^{(2)}-Y^{(2)}_t)
	\end{array}\right) dt
	+
	\left(\begin{array}{cc}
		1 & 0 \\
		0 & \sqrt{Y^{(2)}_t} \\
	\end{array}\right)
	\left(\begin{array}{cc}
		a^{(1,1)} & a^{(1,2)} \\
		a^{(2,1)} & a^{(2,2)} \\
	\end{array}\right)
	\left(\begin{array}{c}
		dW^{(1)}_t \\ dW^{(2)}_t
	\end{array}\right).
\end{equation}
We also assume that there is a risky asset $S^{(1)}$ in the market, that is:
\begin{equation}   \label{eqn:S1}
dS^{(1)}_t = r S^{(1)}_t dt + S^{(1)}_t dR_t,
\end{equation}
where the cumulative excess return $R$ follows the diffusion:
\begin{equation}
dR_t = Y^{(1)}_t \, dt + \sigma \sqrt{Y^{(2)}_t} \, dZ_t, \quad(\sigma\in\mathbb{R}) 
\end{equation}
which is known as the Heston model. In this case the correlation matrix between $Z$ and $W$ is of the form $\rho = (\rho_1, \rho_2)$ satisfying:
\begin{equation}
\left\langle Z, \, W^{(i)} \right\rangle = \rho_i \, dt, \qquad 1 \leq i \leq 2.
\end{equation}

\subsection{Calibration}

Now for the next step we need to set the value of parameters. Let $\mathcal{P}$ be a vector of parameters to be determined given by
\begin{equation}
	\mathcal{P}=\left\{\theta^{(1)}, \theta^{(2)}, k^{(1)}, k^{(2)}, a^{(1,1)}, a^{(1,2)}, a^{(2,1)}, a^{(2,2)}, \sigma, \, \rho_1, \, \rho_2 \right\}.
\end{equation}
We shortly introduce the calibrating process using the nonlinear least squares optimization from the market data. For more detail, see \cite{crisostomo2014analyisis} and \cite{mehrdoust2020calibration}.

Define the Percentage Mean Squared Error (PMSE) between the price $C_{market}$ from the market and the model price $C_{model}$ of the European call option derived from the double Heston model in \cite{mehrdoust2020calibration} and \cite{lemaire2020stationary}:
\begin{equation}
	\text{PMSE}:=\sum_{j=1}^{n} w_j \left(\dfrac{C_{market}(S^{(0)},K_j,T_j,r)-C_{model}(S^{(0)},K_j,T_j,r,\mathcal{P})}{C_{market}(S^{(0)},K_j,T_j,r)} \right)^2,
\end{equation}
where the weights $w_j$ satisfies:
\begin{equation}
	w_j=\dfrac{1}{\sqrt{\left|C_{ask}^{(j)}-C_{bid}^{(j)}\right|}}.
\end{equation}
The optimal parameter vector $\mathcal{P}^\star$ is determined by the following nonlinear least squares problem
\begin{equation}
	\mathcal{P}^\star=\arginf \text{PMSE}.
\end{equation}
\autoref{calib_para} shows the optimal parameters on the observed market data from the S\&P500 index at the close of the market in September 2010.
\begin{table} [h]
	\centering
	\caption{Calibrated parameters}
	\label{calib_para}
	\begin{tabular}{c c c c c c}
	\hline
    $\theta^{(1)}=0.1646$ & $k^{(1)}=0.1301$ & $a^{(1,1)}=-0.6594$ & $a^{(1,2)}=0.7518$ & $\rho_1=-0.2949$ & $\sigma=0.0724$\\
    $\theta^{(2)}=0.2333$ & $k^{(2)}=0.0958$ & $a^{(2,1)}=-0.6692$ & $a^{(2,2)}=0.7431$ & $\rho_2=-0.2919$ & \\ \hline
    \end{tabular}
\end{table}


\subsection{Implementation}

Now let us solve \eqref{eqn:PDE_u} by the DGM algorithm under conditions from the above setting. For the numerical test, we set the interest rate $r=1\%$, the maturity time $T=1$ and the power utility preference parameters $p=0.0005$ and $p=0.5$. We sampled 1000 time-space points $(t,y_1,y_2)$ in the interior of the domain $[0,T]\times[-10,10]\times[0,10]$ and 100 space points at terminal time $T$. We set $100$ steps to resample new time-space domain points. Before resampling, each stochastic gradient descent step is repeated $10$ times. We set $50$ hidden neurons in a hidden layer. From starting $0.001$, learning rate decreased with decay rate 0.96 as the step goes by.
\begin{figure} [h] \centering
	\begin{subfigure}[b]{0.46\textwidth}
		\includegraphics[width=\textwidth]{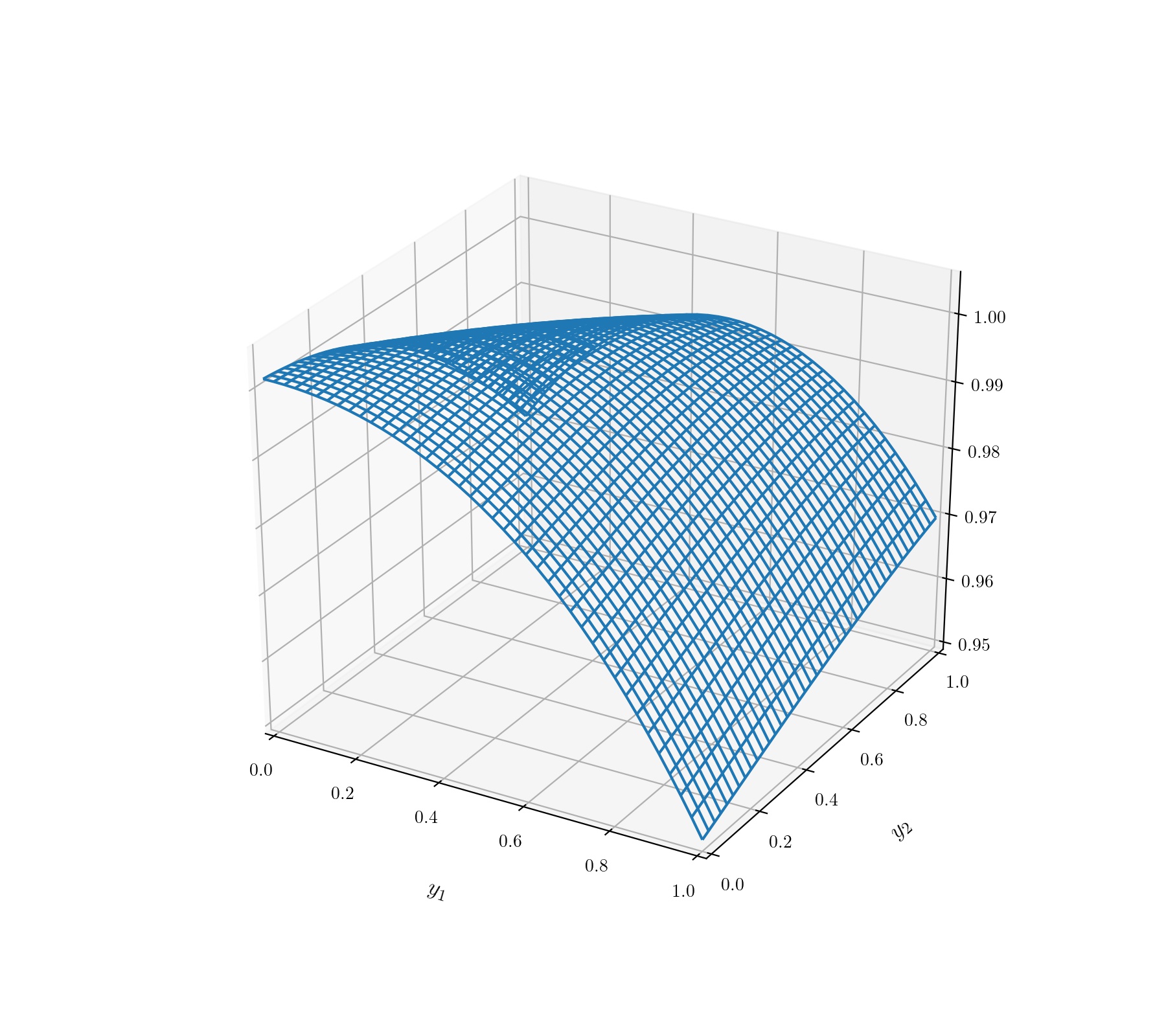}
		\caption{$t=0$}
	\end{subfigure}
	\begin{subfigure}[b]{0.46\textwidth}
		\includegraphics[width=\textwidth]{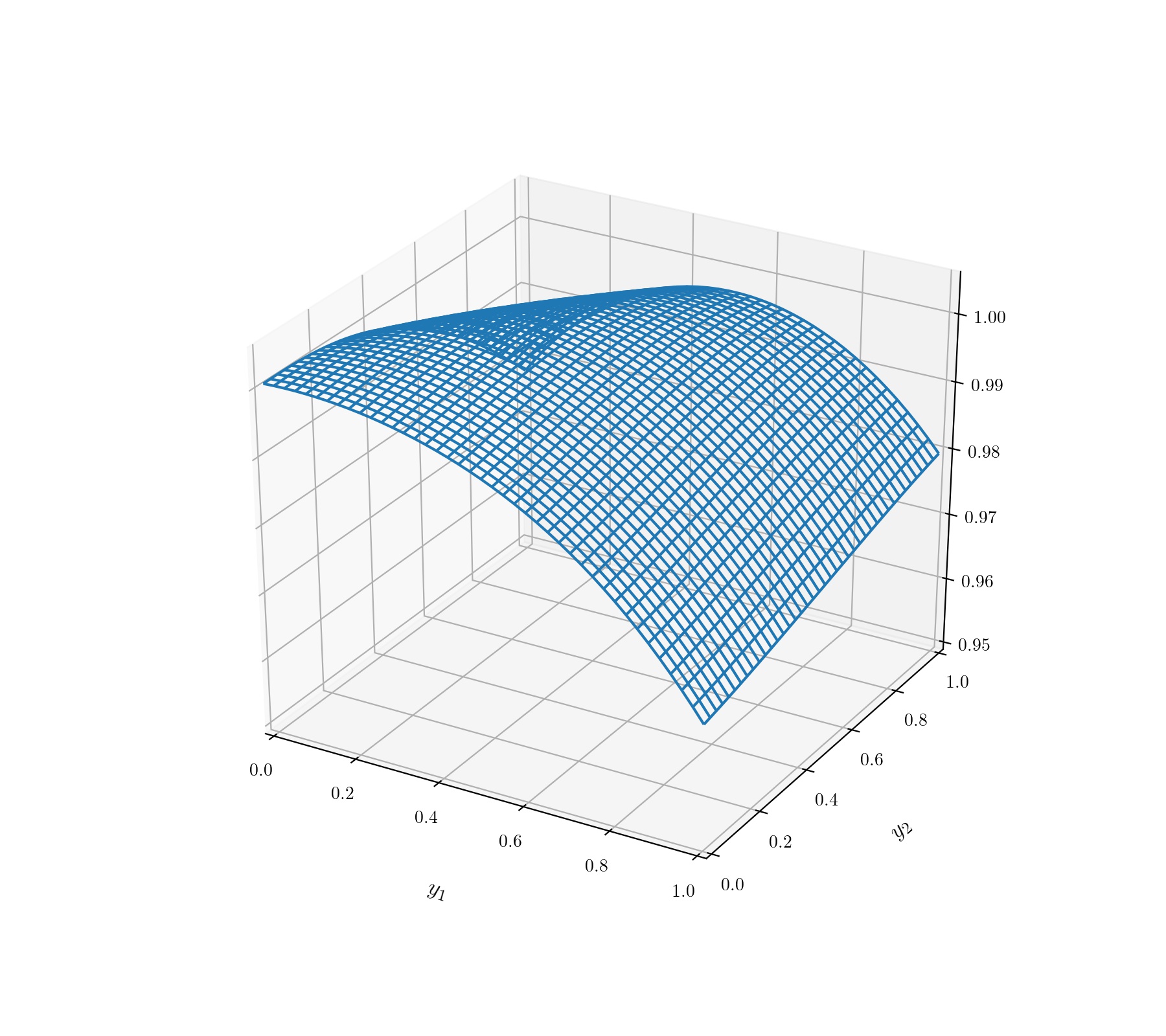}
		\caption{$t=0.25T$}
	\end{subfigure}
	\begin{subfigure}[b]{0.46\textwidth}
    	\includegraphics[width=\textwidth]{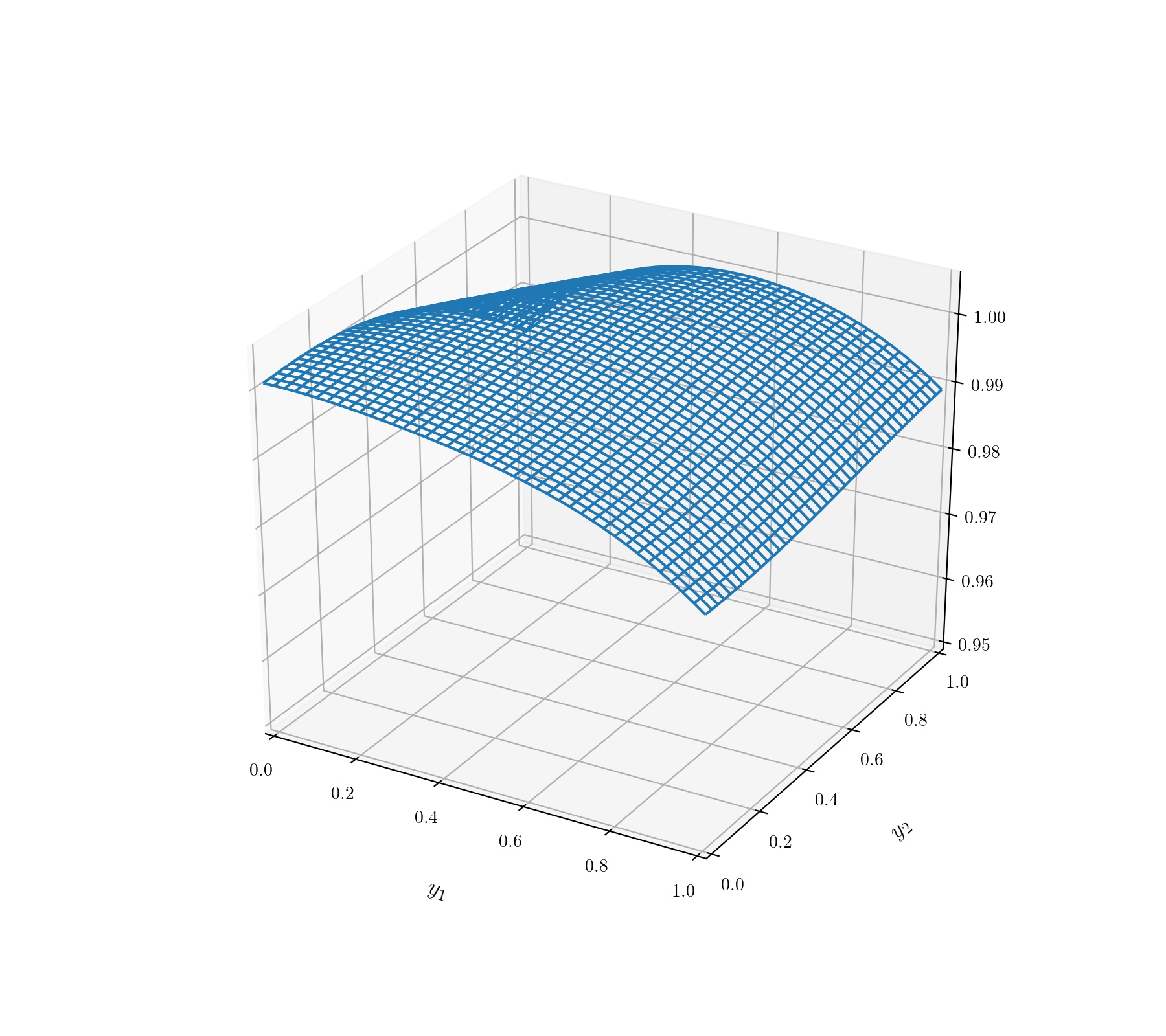}
	    \caption{$t=0.5T$}
    \end{subfigure}
	\begin{subfigure}[b]{0.46\textwidth}
    	\includegraphics[width=\textwidth]{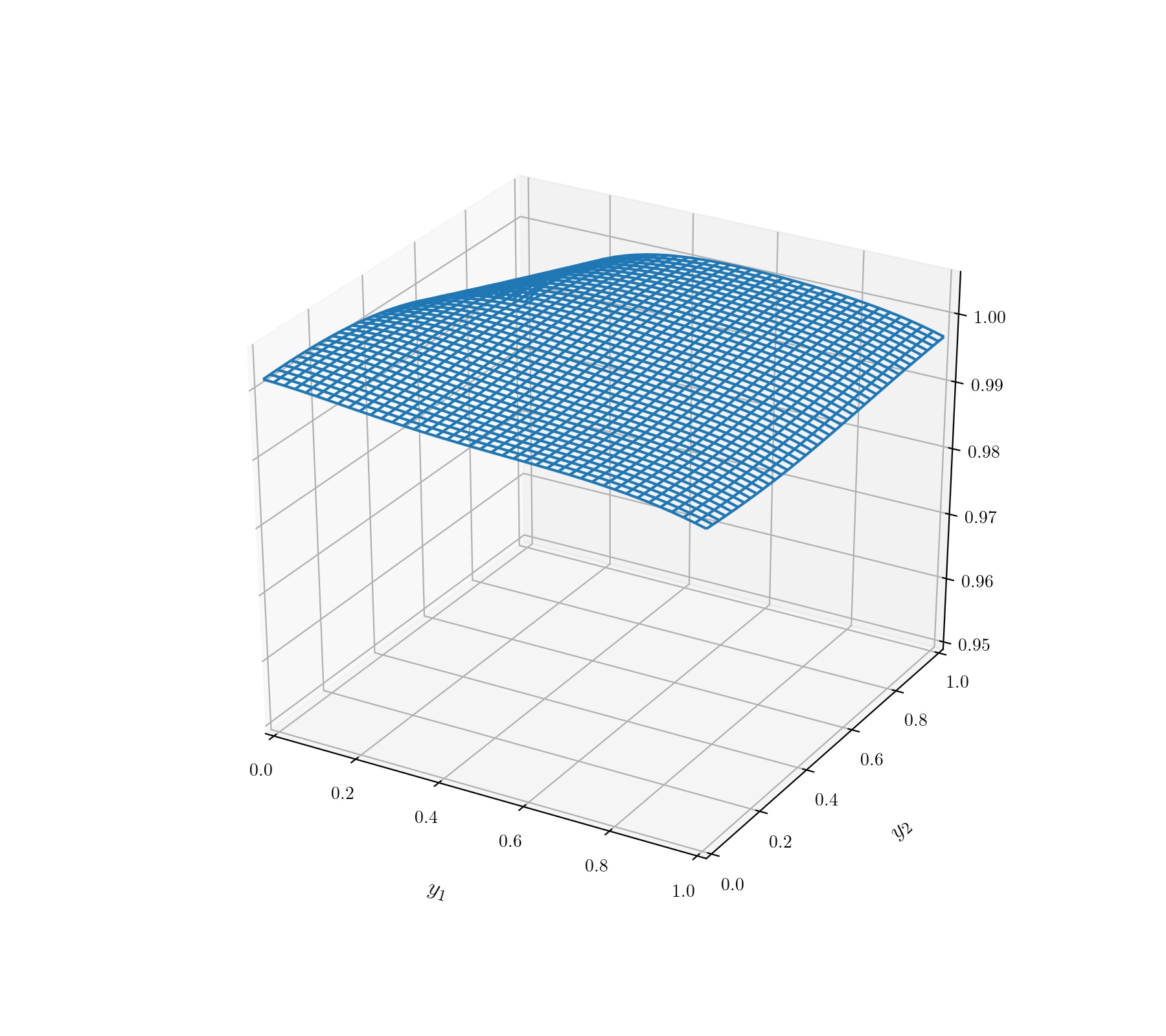}
    	\caption{$t=0.75T$}
    \end{subfigure}
	\caption{Surface of solution by the Deep Galerkin method. $(p=0.0005)$}
	\label{fig:dgm}
\end{figure}

\begin{figure} [h] \centering
	\begin{subfigure}[b]{0.46\textwidth}
		\includegraphics[width=\textwidth]{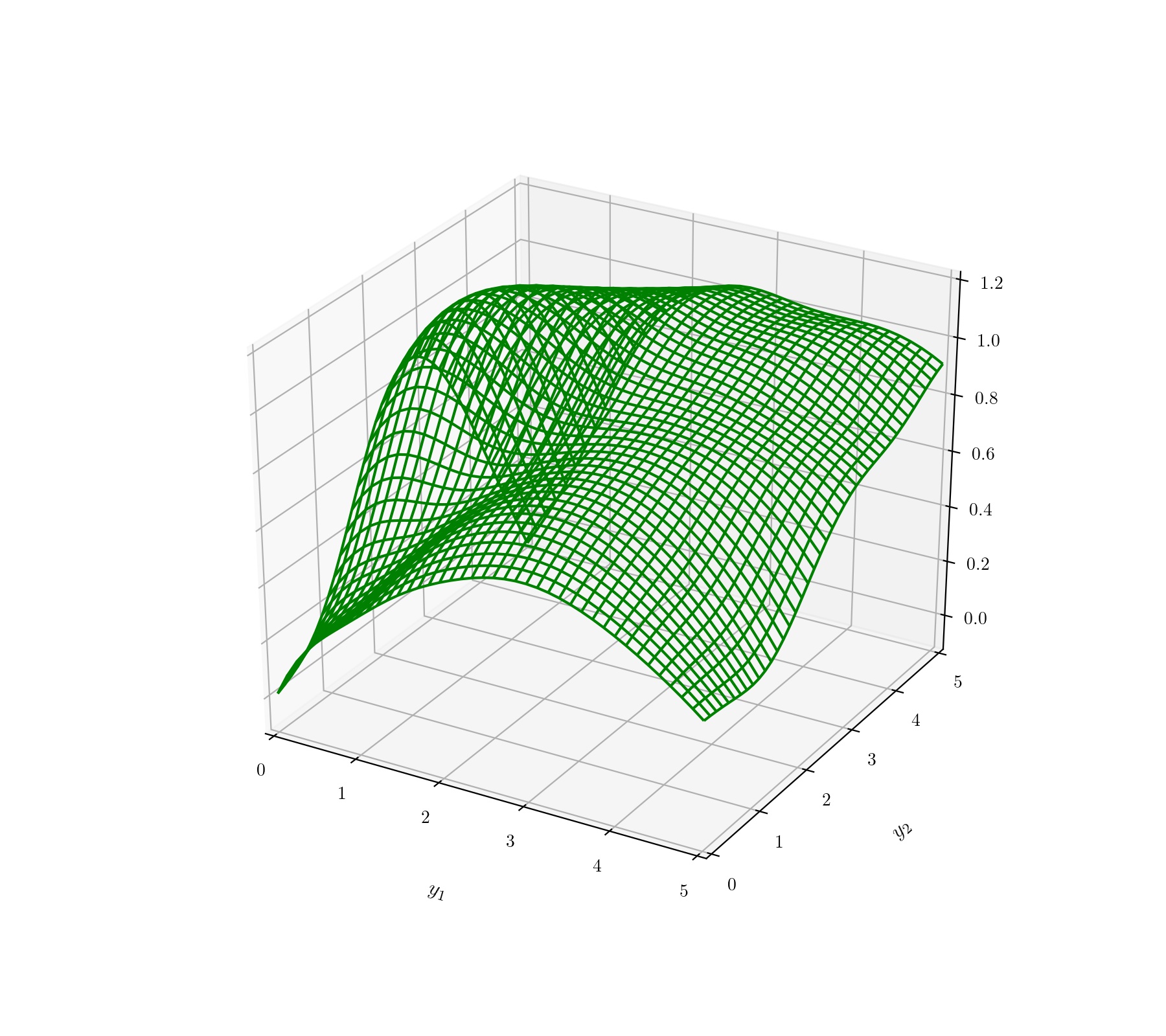}
		\caption{$t=0$}
	\end{subfigure}
	\begin{subfigure}[b]{0.46\textwidth}
		\includegraphics[width=\textwidth]{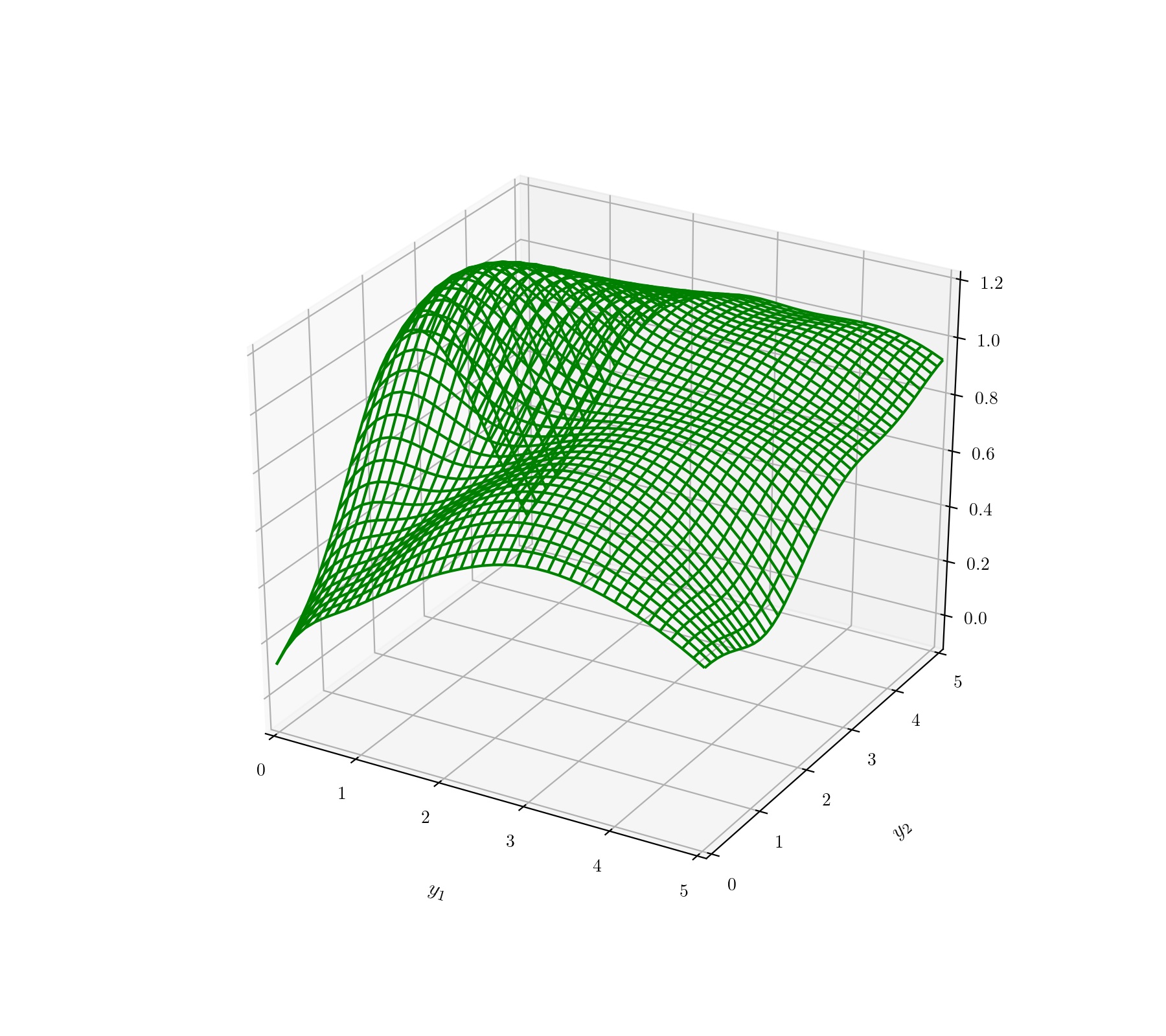}
		\caption{$t=0.25T$}
	\end{subfigure}
	\begin{subfigure}[b]{0.46\textwidth}
		\includegraphics[width=\textwidth]{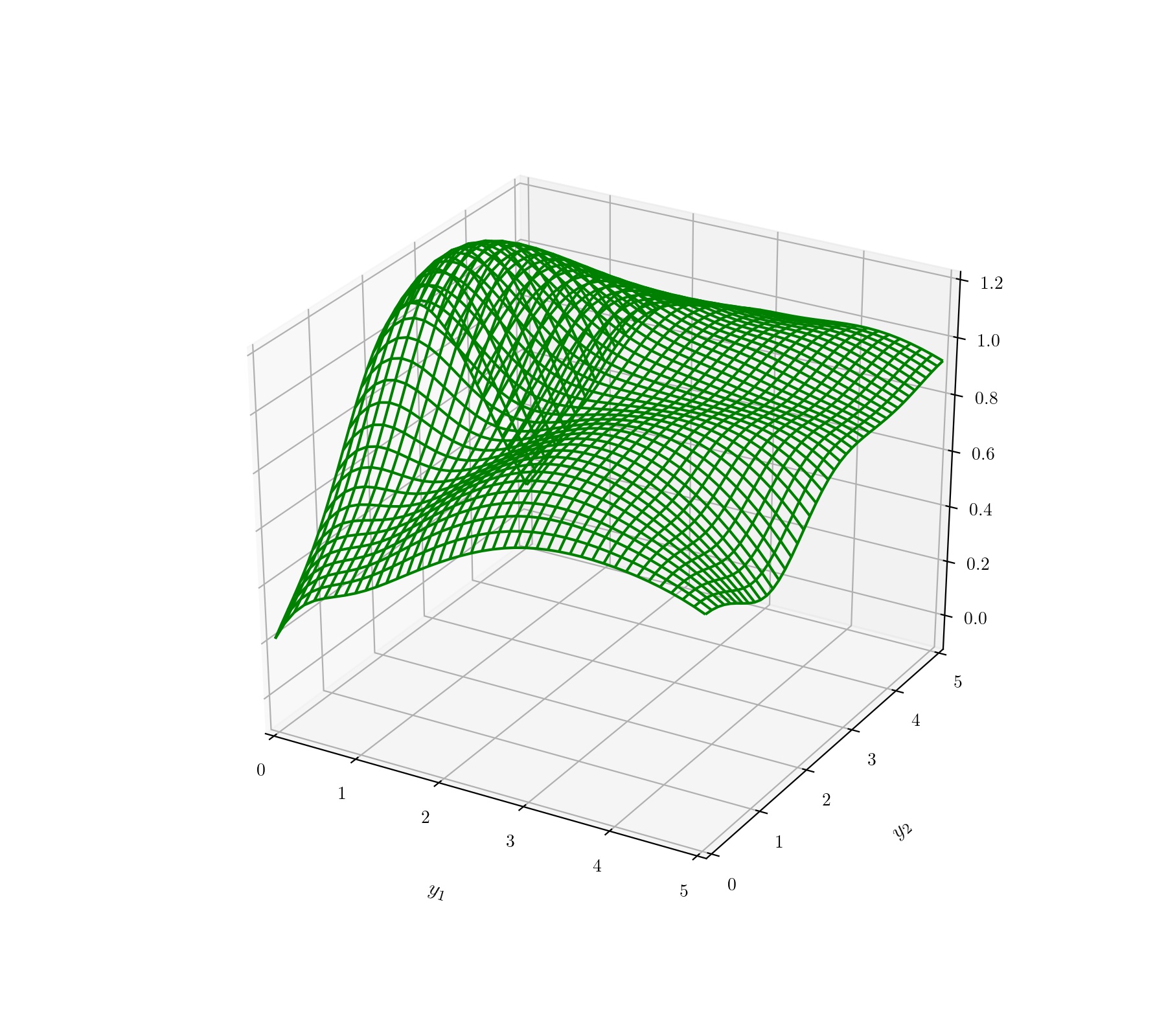}
		\caption{$t=0.5T$}
	\end{subfigure}
	\begin{subfigure}[b]{0.46\textwidth}
		\includegraphics[width=\textwidth]{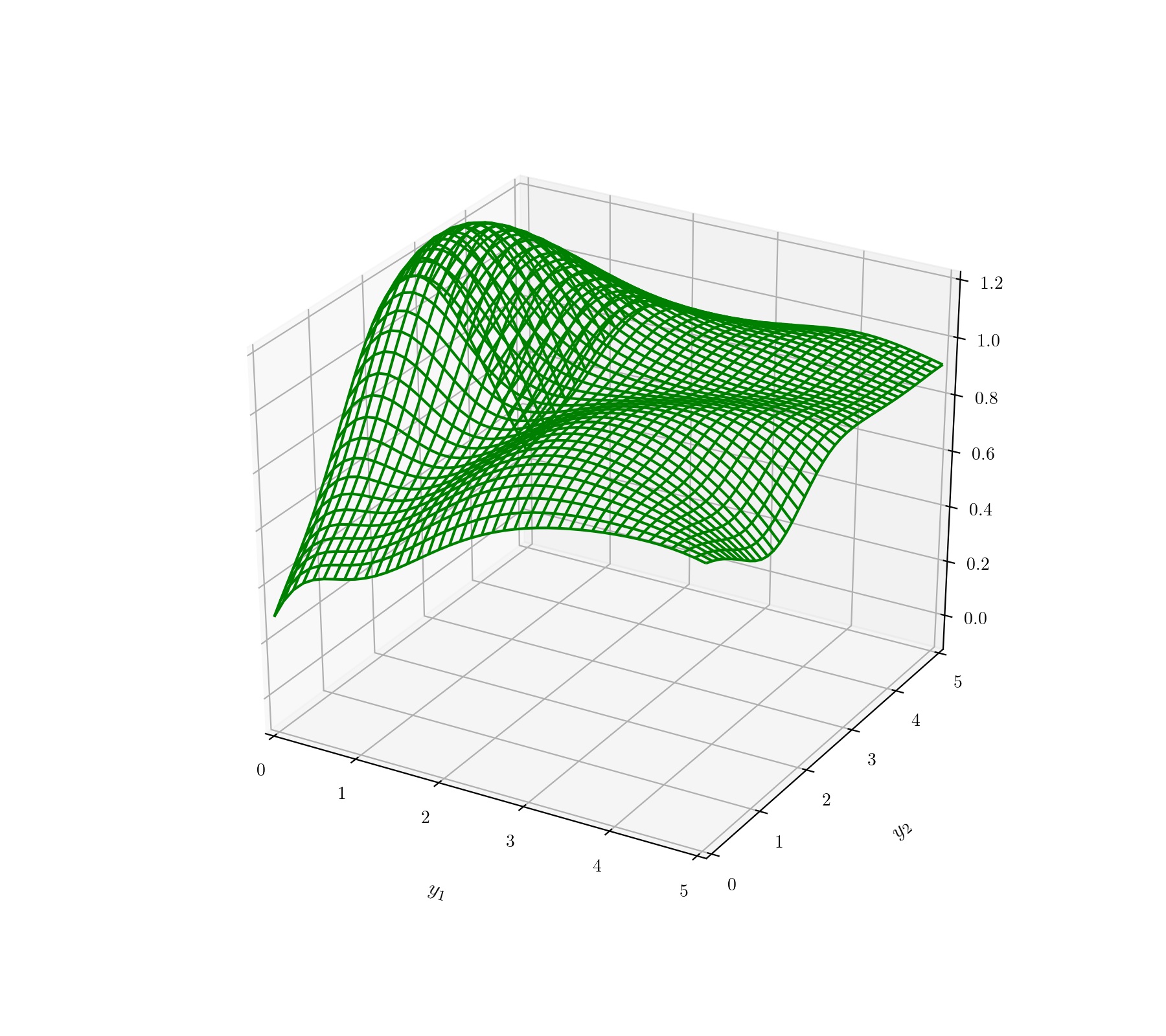}
		\caption{$t=0.75T$}
	\end{subfigure}
	\caption{Surface of solution by the Deep Galerkin method. $(p=0.5)$}
	\label{fig:dgm_0_5}
\end{figure}

After solving \eqref{eqn:PDE_u} by the DGM algorithm, investors can choose their states $(y_1, y_2)\in[-10,10]\times[0,10]$ for fixed $t\in[0,T]$. The optimal portfolio can be constructed using \eqref{eqn:port_u} as:
\begin{equation}
	\pi_{DGM}(t, \, y_1, \, y_2) = \dfrac{1}{1-p} \left( \Sigma^{-1}\mu + \Sigma^{-1}\Upsilon(\nabla_y u_{DGM})\dfrac{1}{u_{DGM}} \right).
\end{equation}
To sum up, one can get the value of $u$ and the portfolio value $\pi$ at every time or state. The investor could buy or sell a risky asset $S^{(1)}$ based on the value of the portfolio to maximize utility from terminal wealth.

\autoref{fig:dgm} shows surfaces of the solution $u_{\text{DGM}}$ of \eqref{eqn:PDE_u} using DGM algorithm in different times, with the power utility preference parameter $p=0.0005$. We chose some part of domain $[0,1]\times[0,1]$ as a plot range for convenience. \autoref{fig:dgm_0_5} shows surfaces of the solution of \eqref{eqn:PDE_u} in $p=0.5$, with the restricted plot range $[0,5]\times[0,5]$. In both figures, for different values of utility parameter $p$, we can easily notice the fact that the surface tends to the plane $u=1$ as time goes to the terminal time $T$: the terminal condition of \eqref{eqn:PDE_u}. Note however \autoref{fig:dgm} is more regular than \autoref{fig:dgm_0_5} in the sense that the value of $L^2$-loss in $p=0.0005$ was remarkably smaller than that in $p=0.5$. Hence we may infer the value of market preference parameter $p$ has played a significant role for using the Deep Galerkin method algorithm.

\subsection{Comparing with the Finite Difference Method}

Now we solve \eqref{eqn:PDE_u} using the finite difference method(FDM). The domain has equally divided $40$ grids satisfying:
\begin{equation}   
	\begin{aligned}
		0&=t^0<t^1<\cdots<t^{40}=T, \\
		-10&=y_1^0<y_1^1<\cdots<y_1^{40}=10, \\
		0&=y_2^0<y_2^1<\cdots<y_2^{40}=10. \\
	\end{aligned}
\end{equation}
First of all, we discretize the solution $u$ as
\begin{equation}
	u_{i,j}^n:=u(t^n,y_1^i,y_2^j), \quad 0 \leq i,j,n \leq 40.
\end{equation}
With this notation, we can substitute the equation \eqref{eqn:PDE_u} using the following central difference formula:
\begin{equation}
	u_t = \dfrac{u_{i,j}^{n+1}-u_{i,j}^{n}}{\Delta t}, \quad u_{y_1} = \dfrac{u_{i+1,j}^{n}-u_{i-1,j}^{n}}{2(\Delta y_1)}, \quad u_{y_2} = \dfrac{u_{i,j+1}^{n}-u_{i,j-1}^{n}}{2(\Delta y_2)}.
\end{equation}
Note that we used the forward difference for discretizing $u_t$ in order to get the values of $(u_{i,j}^n)_{1\leq i,j\leq 40}$ by using the values of $(u_{i,j}^{n+1})_{1\leq i,j\leq 40}$, for $n=0,\ldots,39$.

\begin{figure} [h] \centering
	\begin{subfigure}[b]{0.46\textwidth}
		\includegraphics[width=\textwidth]{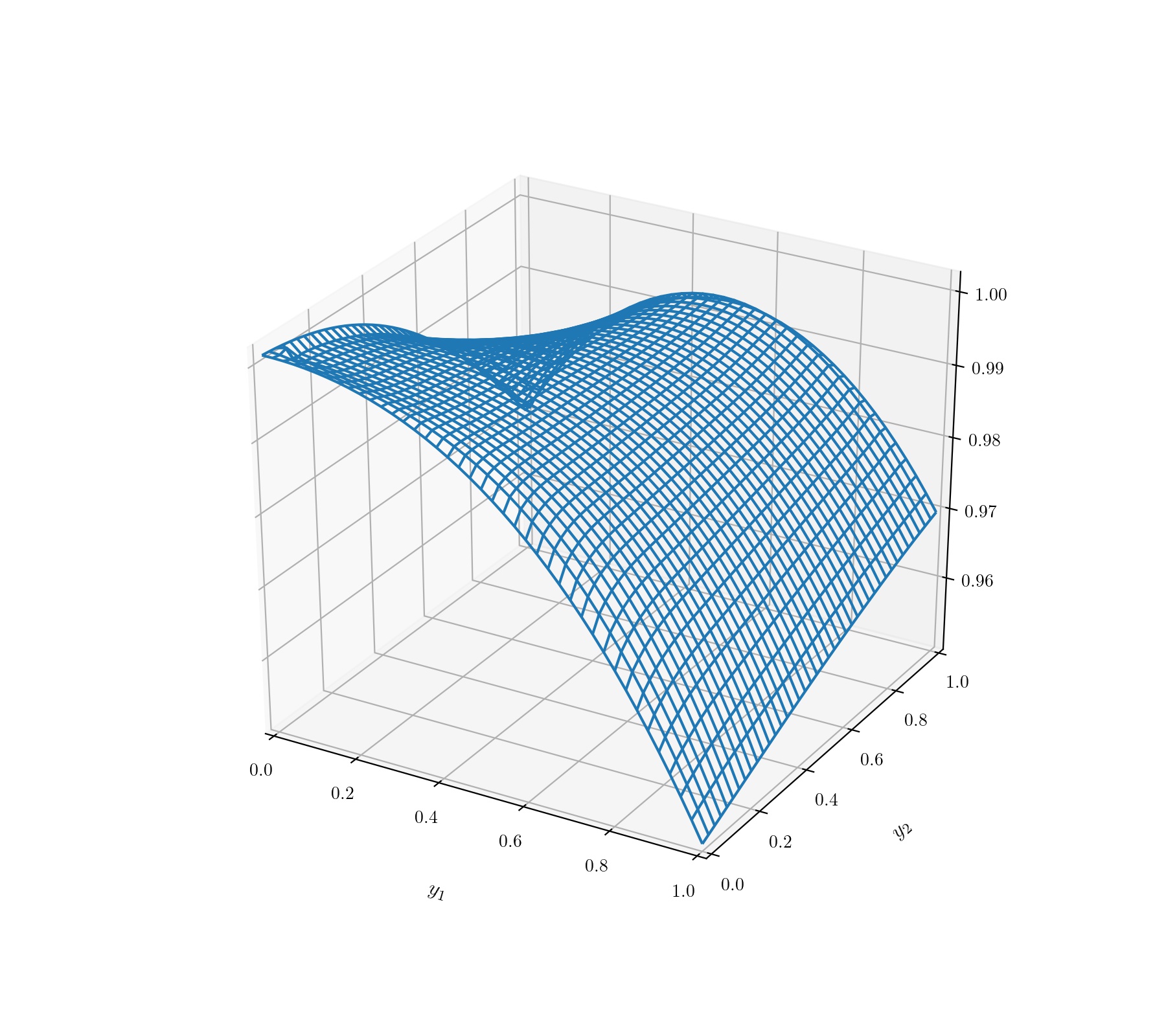}
		\caption{$t=0$}
	\end{subfigure}
	\begin{subfigure}[b]{0.46\textwidth}
		\includegraphics[width=\textwidth]{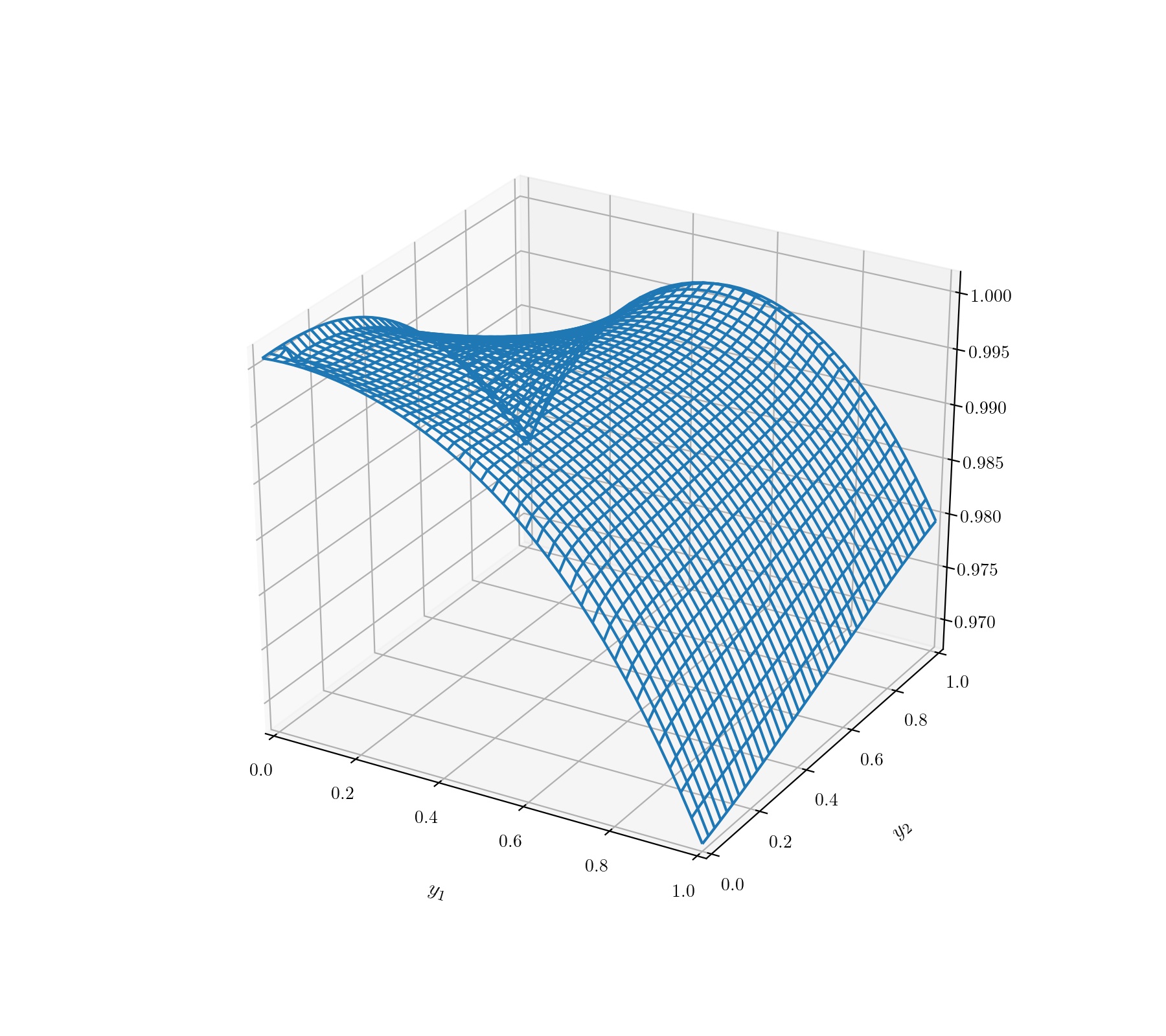}
		\caption{$t=0.25T$}
	\end{subfigure}
	\begin{subfigure}[b]{0.46\textwidth}
		\includegraphics[width=\textwidth]{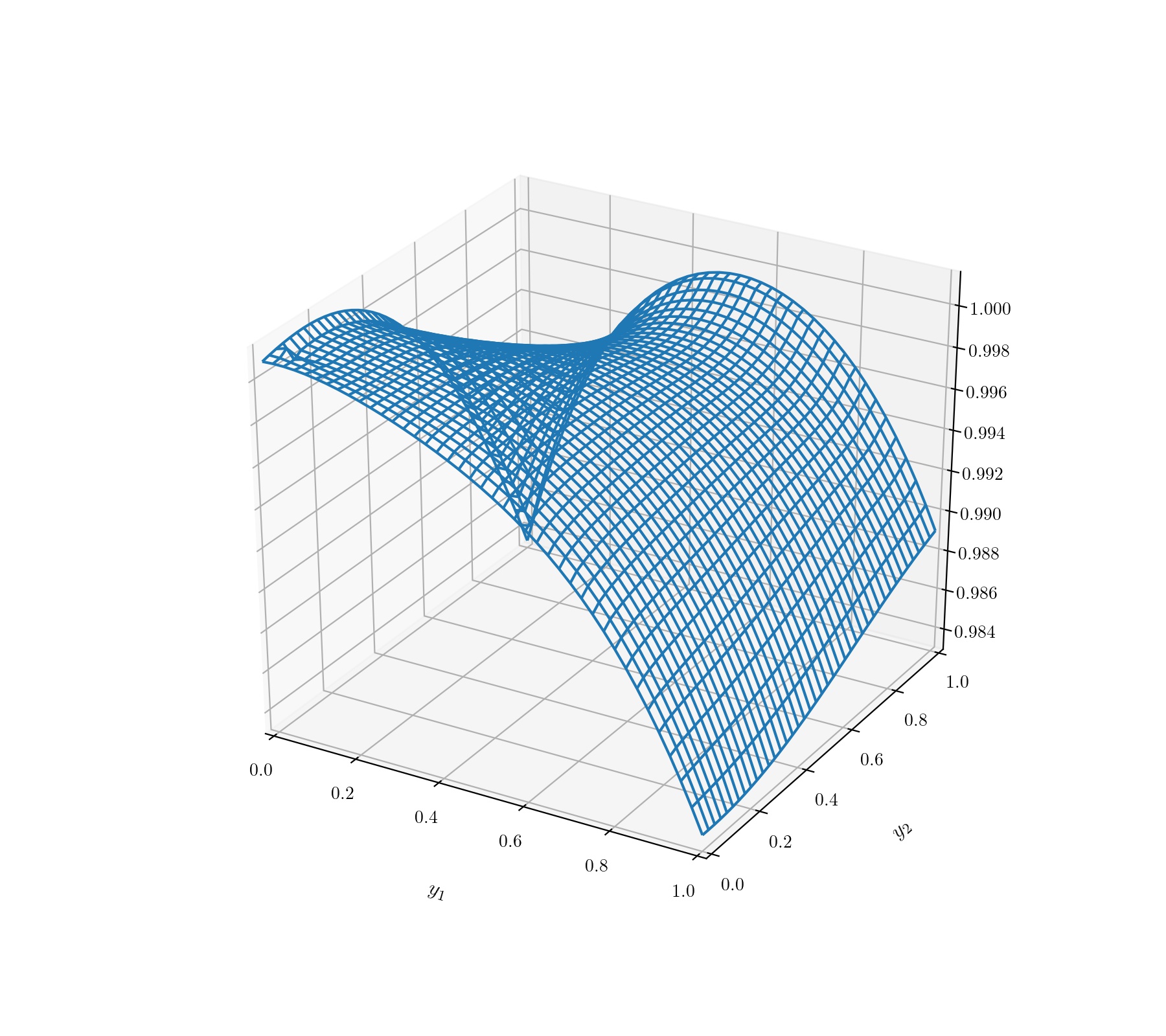}
		\caption{$t=0.5T$}
	\end{subfigure}
	\begin{subfigure}[b]{0.46\textwidth}
		\includegraphics[width=\textwidth]{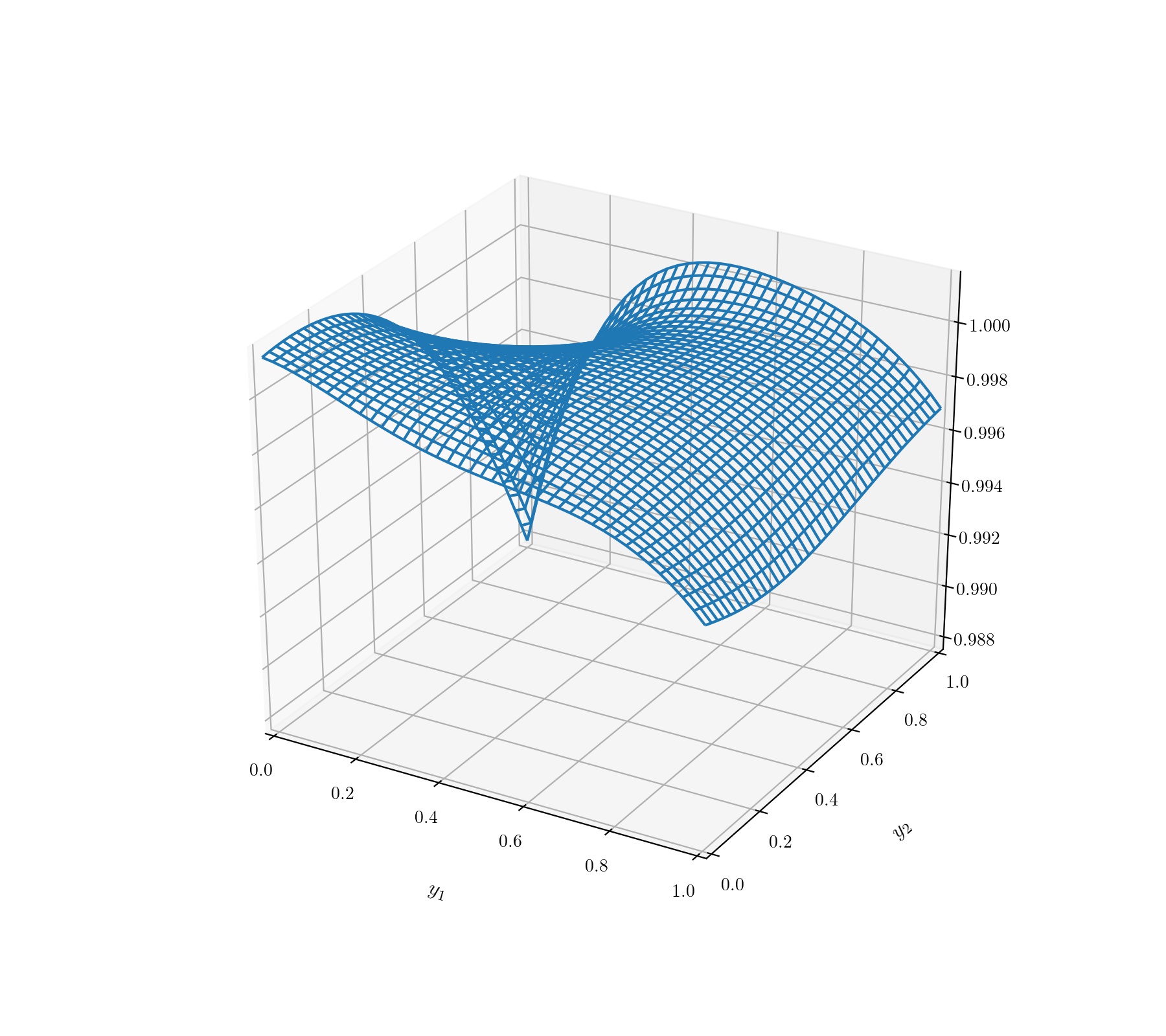}
		\caption{$t=0.75T$}
	\end{subfigure}
	\caption{Surface of solution using the finite difference method}
	\label{fig:fdm}
\end{figure}

Also the central difference approximations of the second derivative of $u$ are given by:
\begin{equation}
	u_{y_1 y_1} = \dfrac{u_{i+1,j}^{n}-2u_{i,j}^{n}+u_{i-1,j}^{n}}{(\Delta y_1)^2}, \quad u_{y_2 y_2} = \dfrac{u_{i,j+1}^{n}-2u_{i,j}^{n}+u_{i,j-1}^{n}}{(\Delta y_2)^2}, 
\end{equation}
\begin{equation}
	u_{y_1 y_2} = \dfrac{u_{i+1,j+1}^{n}-u_{i+1,j-1}^{n}-u_{i-1,j+1}^{n}+u_{i-1,j-1}^{n}}{4(\Delta y_1)(\Delta y_2)}.
\end{equation}
Then the PDE \eqref{eqn:PDE_u} becomes a nonlinear equation with $1521$(=39$\times$39) unknowns $(u^n_{i,j})_{1\leq i,j \leq 39}$ for each $n=0$,$\ldots$,$39$. The equation is of the form:
\begin{equation} \label{eqn:fdm}
\begin{split}
		&\dfrac{u_{i,j}^{n+1}-u_{i,j}^{n}}{\Delta t} + C_1 \, \dfrac{u_{i+1,j}^{n}-u_{i-1,j}^{n}}{2(\Delta y_1)} + C_2 \, \dfrac{u_{i,j+1}^{n}-u_{i,j-1}^{n}}{2(\Delta y_2)} + C_3 \, \dfrac{u_{i+1,j}^{n}-2u_{i,j}^{n}+u_{i-1,j}^{n}}{(\Delta y_1)^2} \\
		&+ C_4 \, \dfrac{u_{i+1,j+1}^{n}-u_{i+1,j-1}^{n}-u_{i-1,j+1}^{n}+u_{i-1,j-1}^{n}}{4(\Delta y_1)(\Delta y_2)} + C_5 \, \dfrac{u_{i,j+1}^{n}-2u_{i,j}^{n}+u_{i,j-1}^{n}}{(\Delta y_2)^2}
		+ C_6 u_{i,j}^{n} \\
		&- \dfrac{q}{2u_{i,j}^{n}}\left[C_7 \left(\dfrac{u_{i+1,j}^{n}-u_{i-1,j}^{n}}{2(\Delta y_1)}\right)^2 + C_8 \dfrac{u_{i+1,j}^{n}-u_{i-1,j}^{n}}{2(\Delta y_1)} \dfrac{u_{i,j+1}^{n}-u_{i,j-1}^{n}}{2(\Delta y_2)} + C_9 (\dfrac{u_{i,j+1}^{n}-u_{i,j-1}^{n}}{2(\Delta y_2)})^2\right]=0,
\end{split}
\end{equation}
where $C_1,\ldots,C_9$ are constants.
Note that the terminal condition $u(T,y)=u(T,y_1,y_2)=1$ also becomes
\begin{equation}
	u^{40}_{i,j}=1 \quad \text{for all} \quad 0 \leq i,j \leq 40.
\end{equation}
Since \eqref{eqn:PDE_u} has no boundary condition, we used the boundary data from the DGM algorithm. \autoref{fig:fdm} shows surfaces of the solution of \eqref{eqn:fdm} using the finite difference method in different times with $p=0.0005$. We used the Newton's method since the equation \eqref{eqn:fdm} is nonlinear. For more detail, see \cite{remani2013numerical}.

\begin{figure} [h] \centering
	\begin{subfigure}[b]{0.46\textwidth}
		\includegraphics[width=\textwidth]{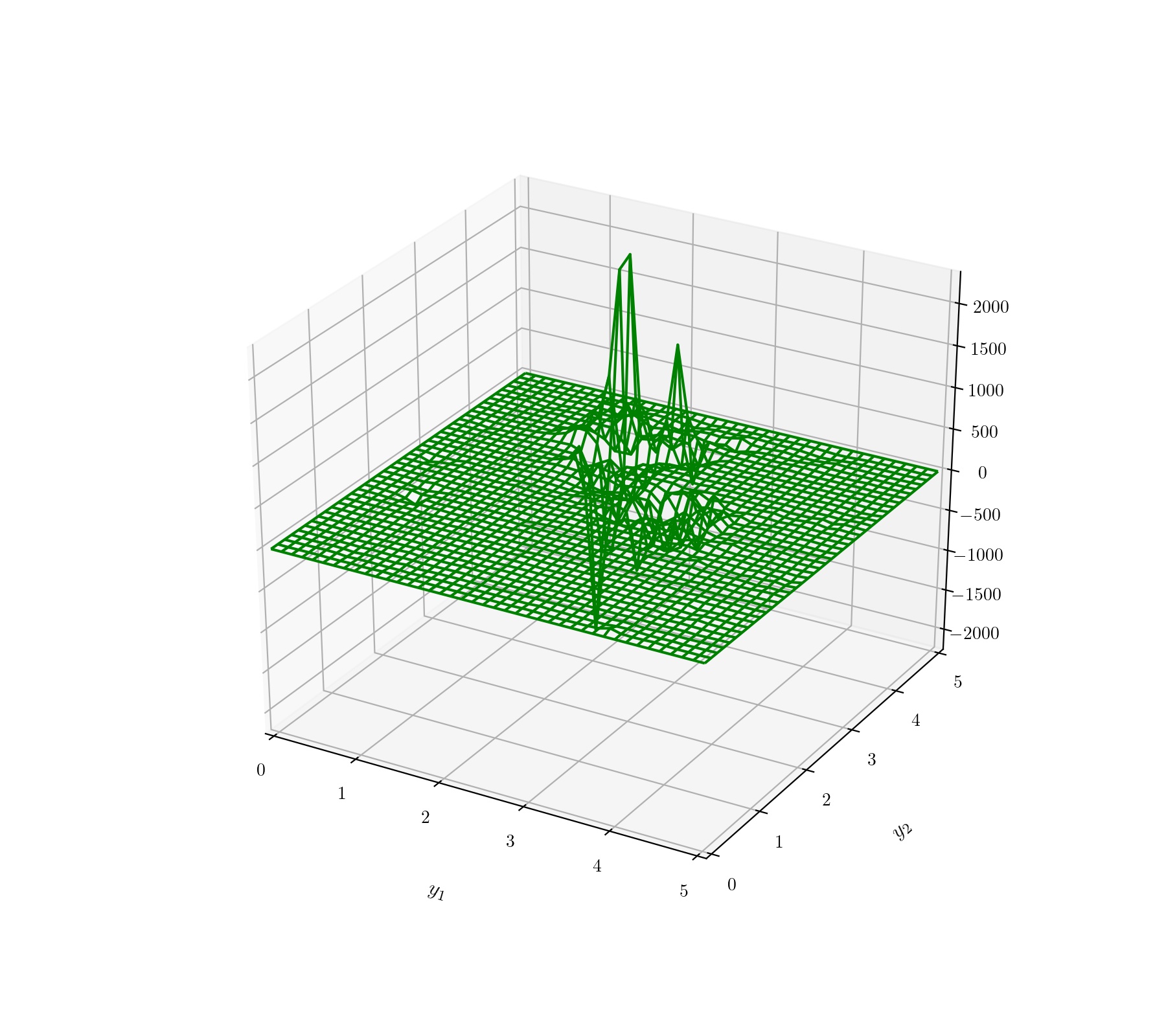}
		\caption{$t=0$}
	\end{subfigure}
	\begin{subfigure}[b]{0.46\textwidth}
		\includegraphics[width=\textwidth]{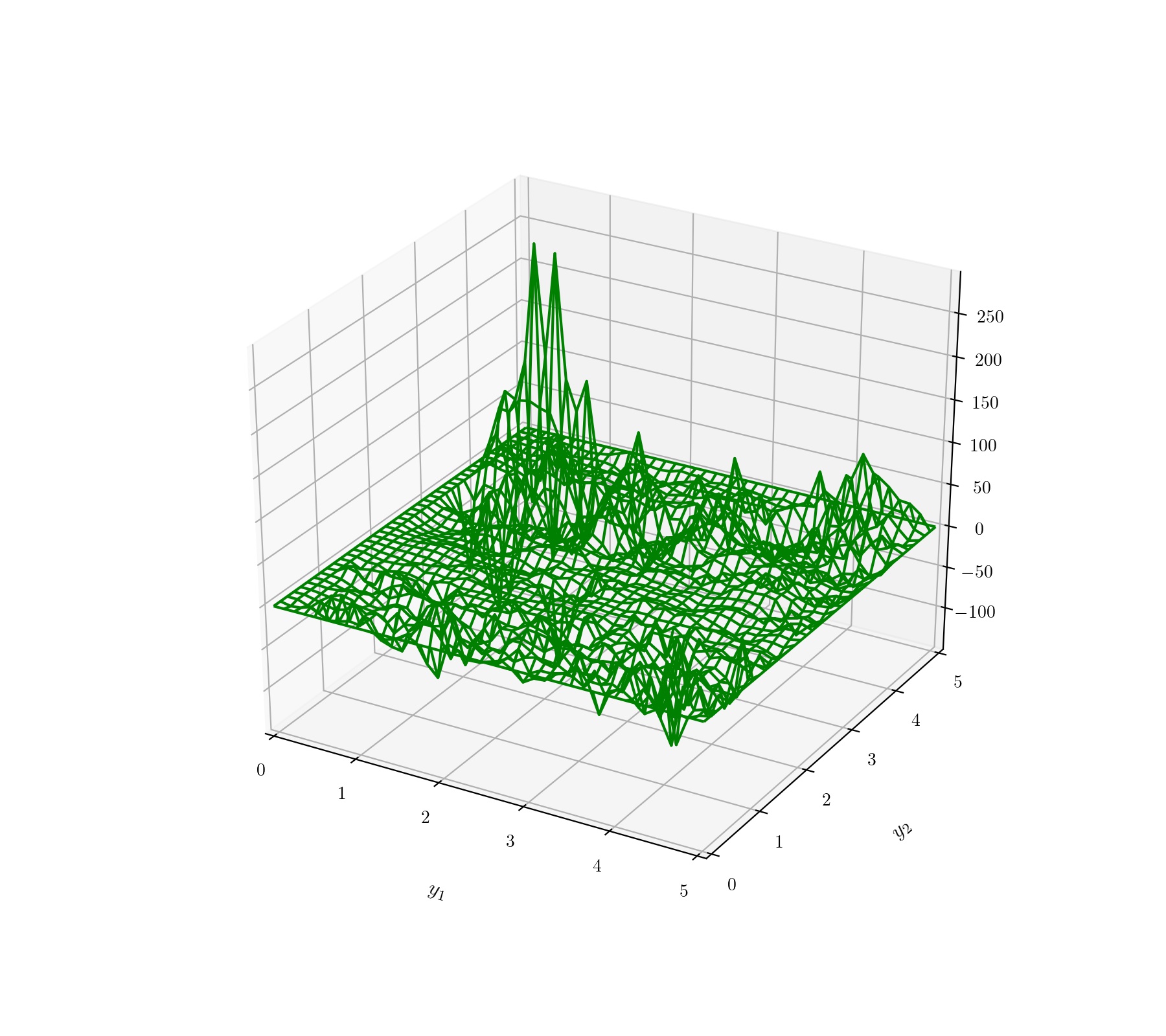}
		\caption{$t=0.25T$}
	\end{subfigure}
	\begin{subfigure}[b]{0.46\textwidth}
		\includegraphics[width=\textwidth]{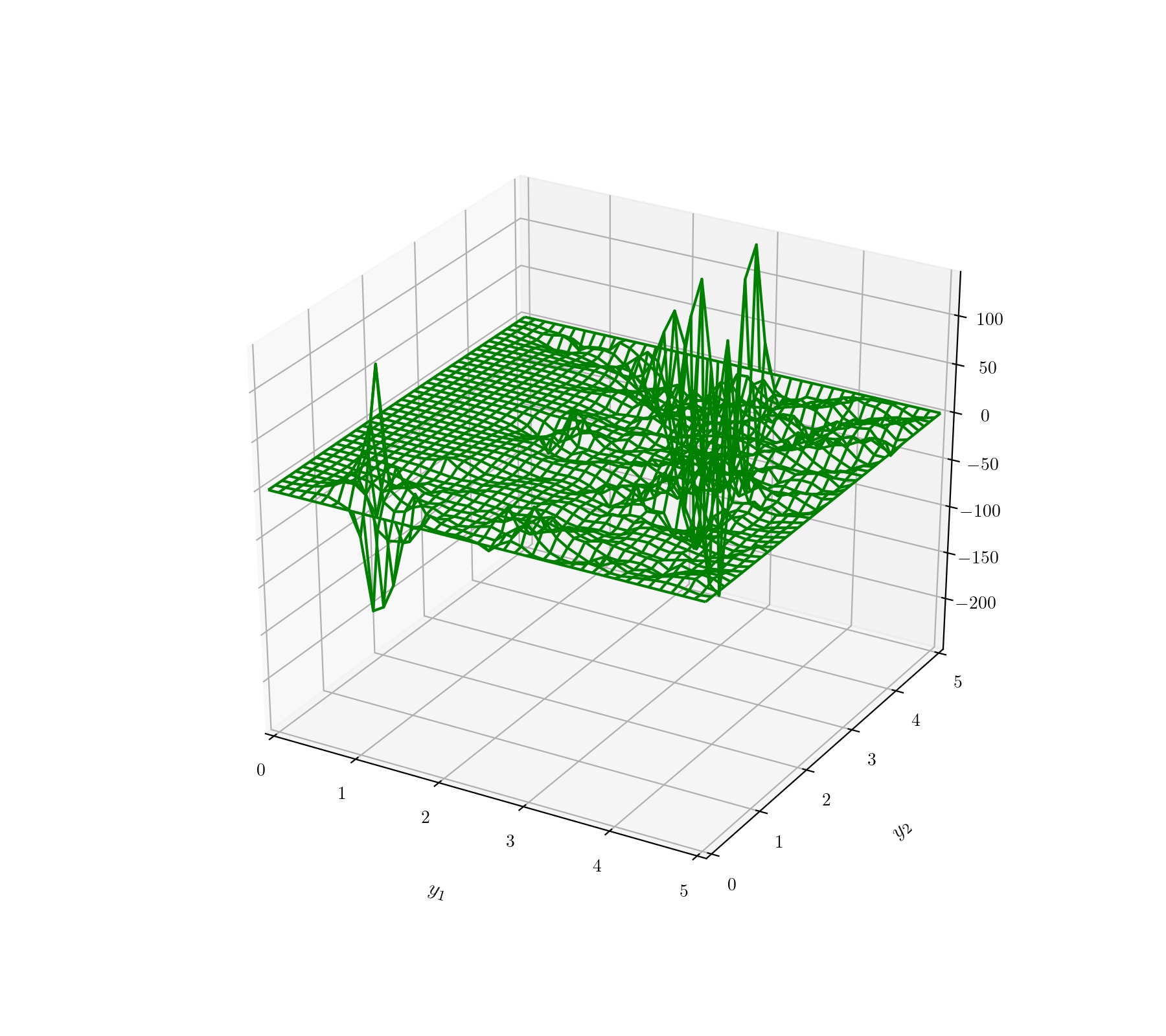}
		\caption{$t=0.5T$}
	\end{subfigure}
	\begin{subfigure}[b]{0.46\textwidth}
		\includegraphics[width=\textwidth]{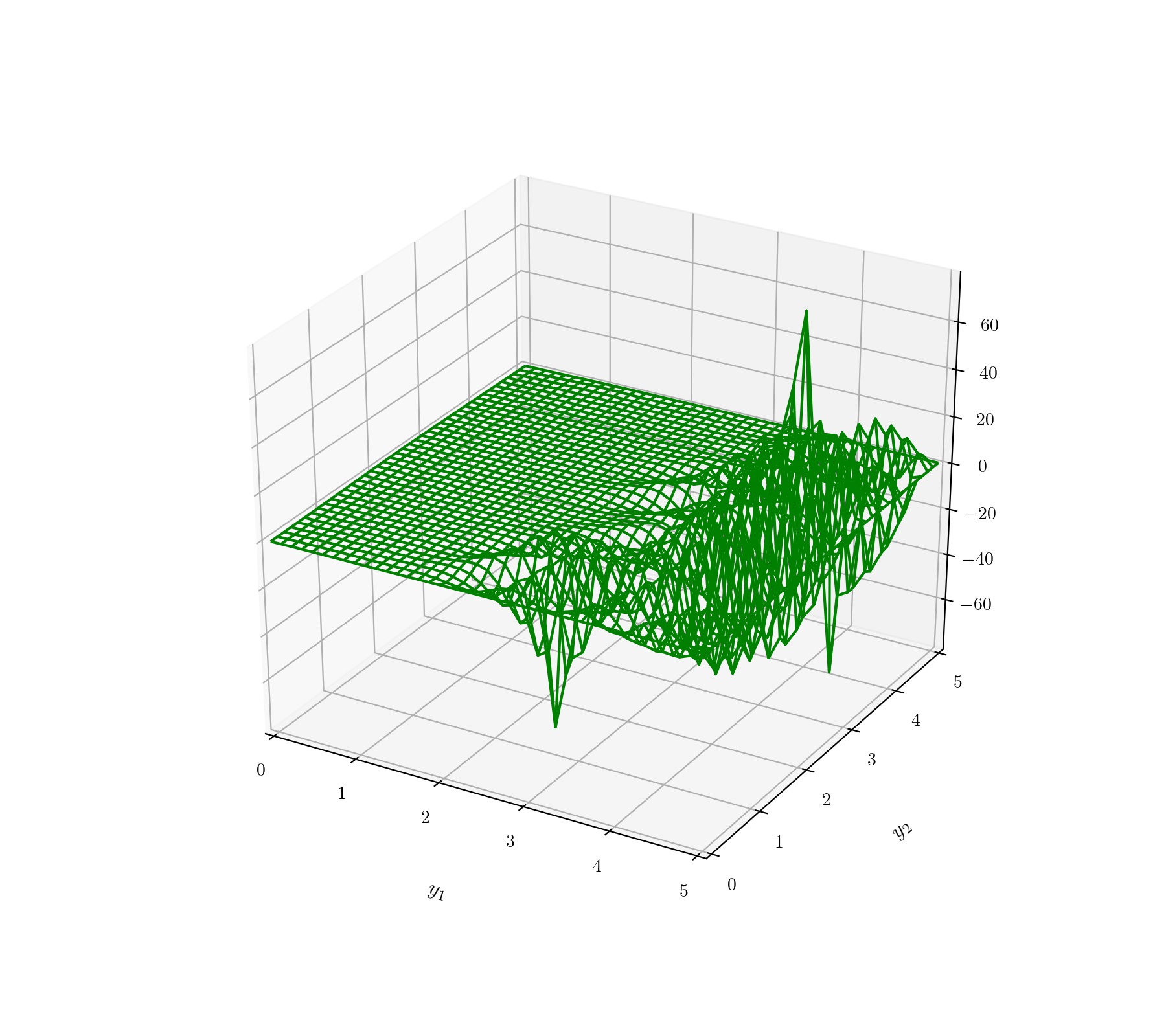}
		\caption{$t=0.75T$}
	\end{subfigure}
	\caption{Surface of solution by the finite difference method $(p=0.5)$}
	\label{fig:fdm_0_5}
\end{figure}

\begin{figure} [!t] \centering
	\begin{subfigure}[b]{0.36\textwidth}
		\includegraphics[width=\textwidth]{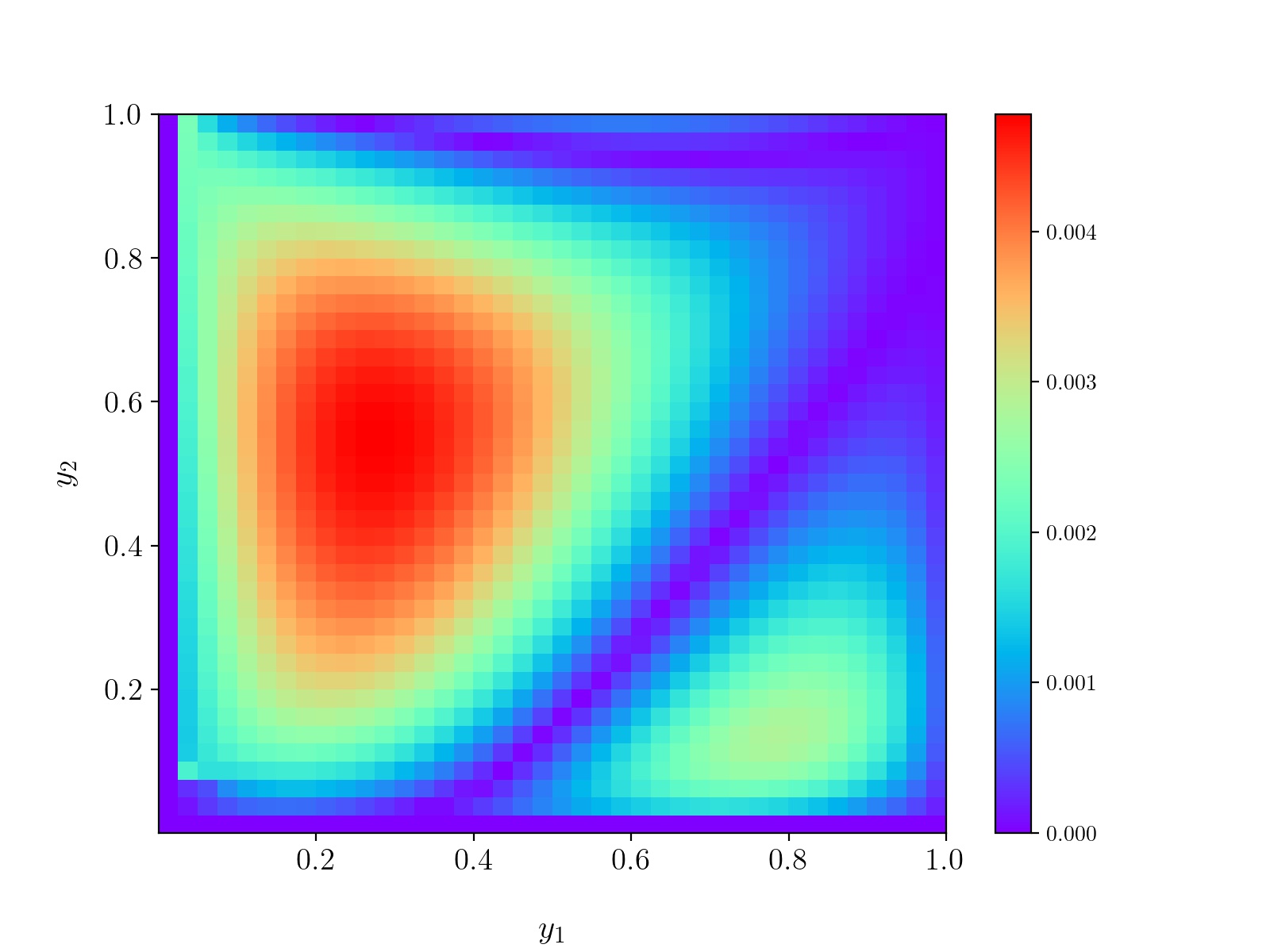}
		\caption{$t=0$}
	\end{subfigure}
	\begin{subfigure}[b]{0.36\textwidth}
		\includegraphics[width=\textwidth]{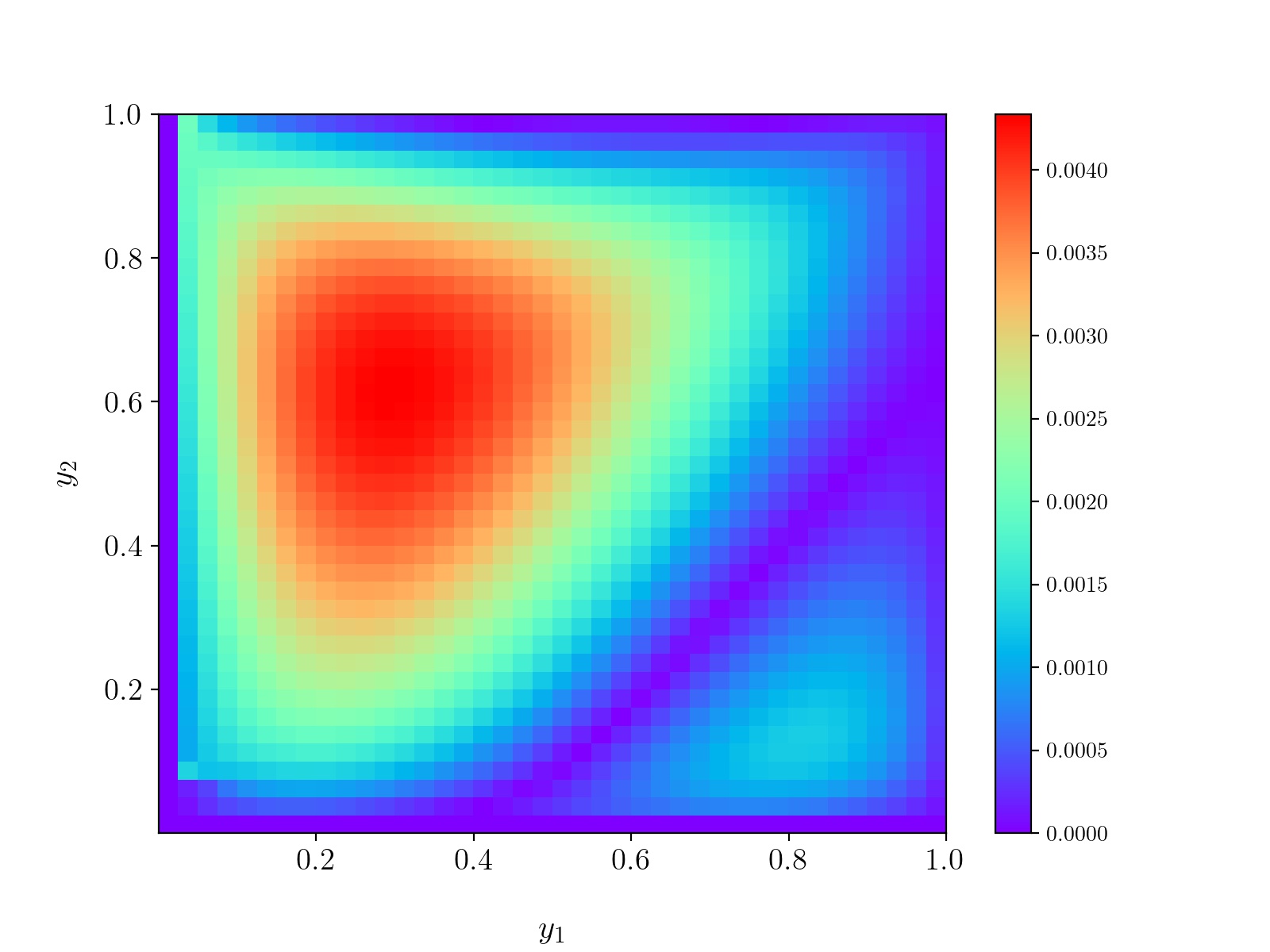}
		\caption{$t=0.25T$}
	\end{subfigure}
	\begin{subfigure}[b]{0.36\textwidth}
		\includegraphics[width=\textwidth]{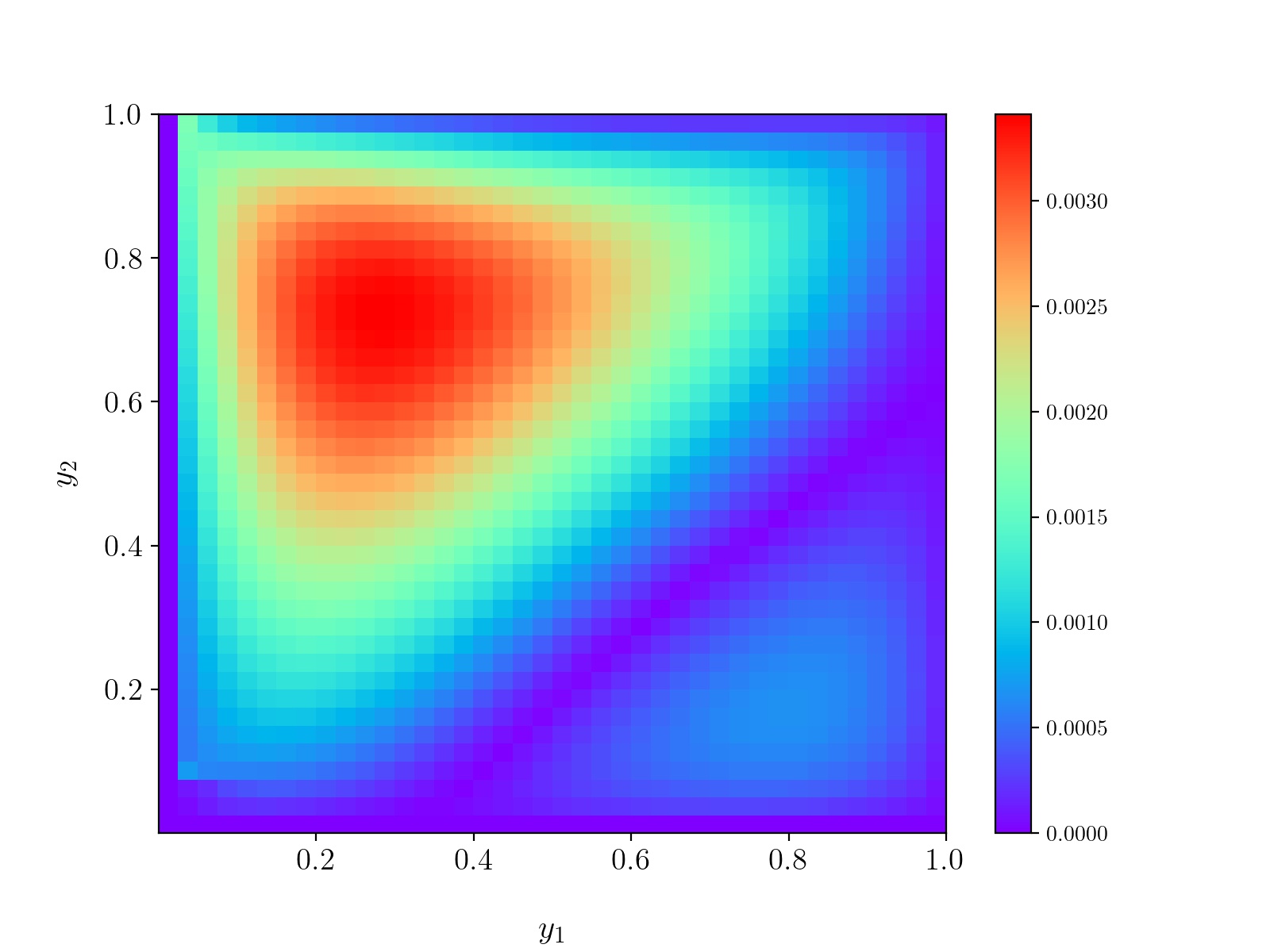}
		\caption{$t=0.5T$}
	\end{subfigure}
	\begin{subfigure}[b]{0.36\textwidth}
		\includegraphics[width=\textwidth]{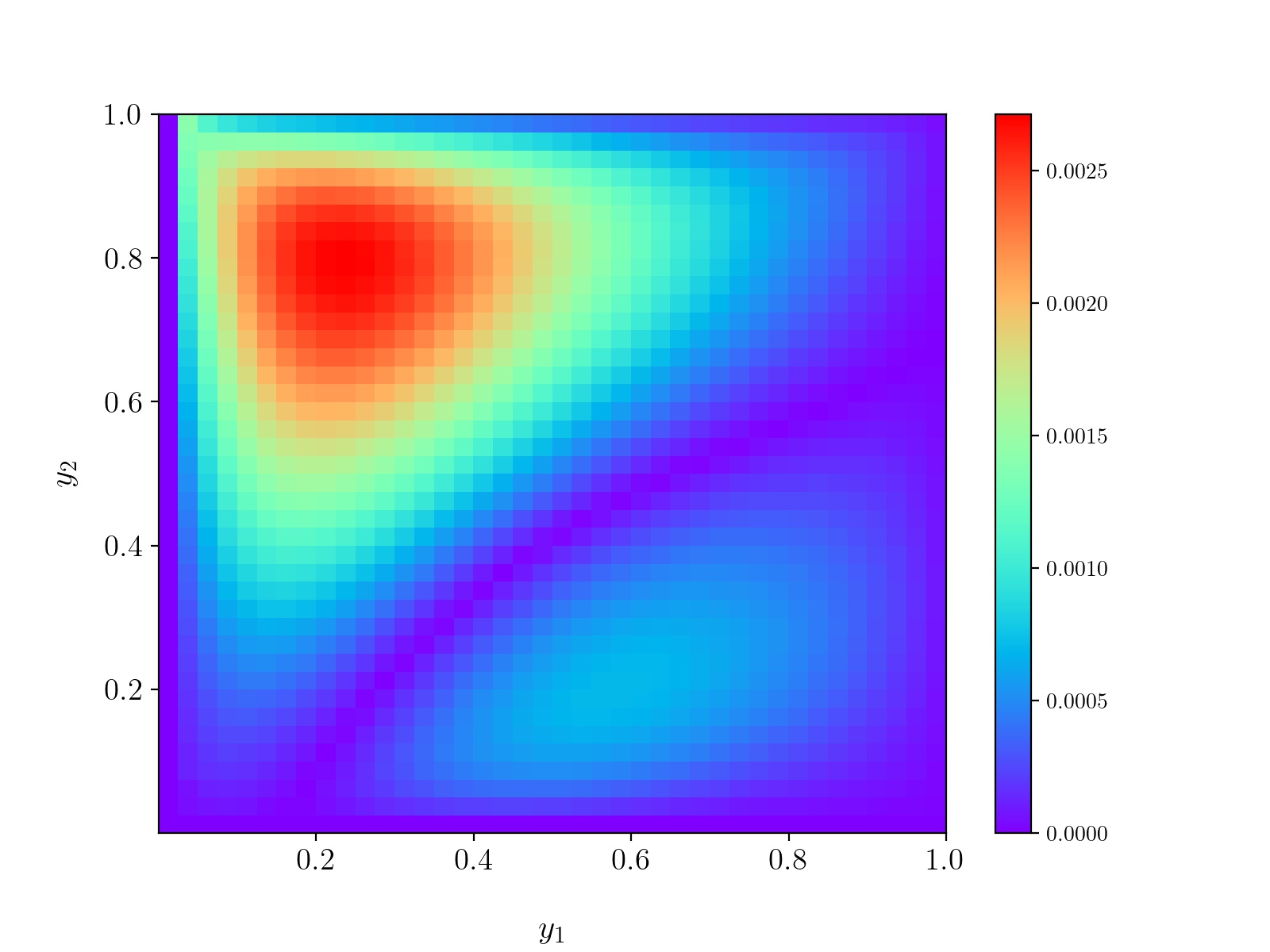}
		\caption{$t=0.75T$}
	\end{subfigure}
	\caption{Absolute errors between the Deep Galerkin method and the finite difference method}
	\label{fig:abs_error}
\end{figure}

With the same value of $p$, the absolute errors between the solution from the Deep Galerkin method and the one from the finite difference method are displayed in \autoref{fig:abs_error}. Notice that the error between these algorithms is getting slightly larger as the time $t$ goes to zero. This may be due to the time-reversely performed finite difference method algorithm, from $t=T$ to $t=0$. In other words, the stability on the solution from the terminal condition was gradually weakened as the time goes to zero.

In a different point of view, combining \autoref{fig:abs_error} with \autoref{fig:dgm} and \autoref{fig:fdm}, we conclude the solution is well-estimated by the deep neural network. It usually takes about $5$ minutes to train the network. On the other hand, it only takes less than $30$ seconds to find the surface of solution by the FDM. One can deduce this traditional algorithm would be more efficient for time-saving. However, it is not always true. \autoref{fig:fdm_0_5} shows surfaces derived from the finite difference method algorithm with $p=0.5$, same domain with \autoref{fig:dgm_0_5}. In \autoref{fig:fdm_0_5}, the solution has extremely large or small values. This singularity may have arised since the system of equations \eqref{eqn:fdm} is nonlinear. In other words, the matter of finding inverse matrix in the Newton's method at each step $n=39,\ldots,0$ would make the value of solutions undesirable.

\section{Conclusion} \label{sec5}

In this paper we first modeled the market with a safe asset and some risky assets whose dynamics satisfy the diffusion process with returns. We then induced the HJB equation to maximize the expectation of an investor's utility, given by investment opportunities modeled by a $d$-dimensional state process. Using some properties including homotheticity and concaveness, we finally derived a nonlinear partial differential equation and approximated the solution with a deep learning algorithm.

For comparison with the Deep Galerkin method, we applied the finite difference method to find an approximated solution. In case of the utility parameter being quite small, $p=0.0005$, we found that the solution has estimated well by the neural network. However in the case of $p=0.5$, there were several singular points in solution surfaces approximated by the finite difference method. Hence unlike the Deep Galerkin method, this mesh-based algorithm showed some defects such as a singularity by a nonlinearity of discretized version of partial differential equations. This concludes that the DGM algorithm is relatively stable and has less difficulties to approximate the solution for PDEs.

Furthermore, all above procedures in \autoref{sec4} were performed only with the $2$-dimensional state process. If the dimension $d$ of state process increases, since there would exist millions of grids, it would be more computationally efficient to apply the DGM algorithm than the FDM algorithm. Finally with the approximated solution from the relatively stable DGM algorithm, the investor can decide how to allocate one's wealth in several risky assets by the optimal portfolio formula.

Also there has some further studies to be researched. the stability or regularity of the solution is to be researched as the following are changed: model or dimension of a state variable $Y$, value of calibrated parameters, market preference parameter $p$ and sampling domain. Also in the optimal portfolio formula, the stability on a gradient term needs to be considered. Meanwhile, \cite{sirignano2018dgm} proved the convergence of the DGM algorithm only in a class of quasilinear parabolic PDEs. Although \cite{sirignano2018dgm} refered that the algorithm can be applied to other types of PDEs, there needs to be some researches for the stability of hyperbolic, elliptic or fully nonlinear PDEs.

\begin{appendices} \label{appendix}

\section{Proof of Theorem \ref{thm:DGM}}
Here we now justify Theorem \ref{thm:DGM} by proving the following two theorems in special cases. The main idea of proofs are from \cite{sirignano2018dgm} and \cite{hornik1991approximation} based on universal approximation arguments. Note that the formulations in this section are not the same as the ones from the above papers. For completeness, we display almost all computations in the following proofs. The first theorem shows the convergence of $J(f)$: there exists a deep neural network $f$ such that the loss functional $J(f)$ tends to the arbitrary small. The latter one stands for the convergence of the DNN function to the solution of PDEs.
\subsection{Convergence of the loss functional}
Assume $D\subset\mathbb{R}^d$ is bounded with a smooth boundary $\partial D$. Denote $D_T =[0,T)\times D$. Consider the following form of quasilinear parabolic PDE:
\begin{equation}   
	\begin{aligned}
		\mathcal{G}[u](t,y):=\partial_t u(t,y)-\text{div}(\alpha(t,y,u,\nabla u))+\gamma(t,y,u,\nabla u)&=0, &(t,y) \in D_T, \\
		u(T,y)&=u_T(y), &y \in D. \label{eqn:para_pde} \\
	\end{aligned}
\end{equation}
Then the above differential operator $\mathcal{G}$ can be expressed as
\begin{equation}
	\begin{split}
		\mathcal{G}[u](t,y)&=\partial_t u(t,y)-\sum_{i,j=1}^{d}\dfrac{\partial\alpha_i}{\partial u_{y_j}}\dfrac{\partial u_{y_j}}{\partial y_i}-\sum_{i=1}^{d}\dfrac{\partial\alpha_i}{\partial u}\partial_{y_i}u-\sum_{i=1}^{d}\dfrac{\partial\alpha_i}{\partial_{y_i}}+\gamma(t,y,u,\nabla u) \\
		&=:\partial_t u(t,y)-\sum_{i,j=1}^{d}\dfrac{\partial\alpha_i}{\partial u_{y_j}}\dfrac{\partial u_{y_j}}{\partial y_i}+\hat{\gamma}(t,y,u,\nabla u).
	\end{split}
\end{equation}
\begin{thm} \label{thm_no_1}
	Let $\mathfrak{C}^n(\psi)$ be a collection of DNN functions with $n$ hidden neurons in a single hidden layer:
	\begin{equation}
		\mathfrak{C}^n(\psi)=\left\{\zeta:\mathbb{R}^{1+d} \rightarrow \mathbb{R} : \zeta(t,y)=\sum_{i=1}^{n}\beta_i\psi\left(\alpha_{1i}t+\sum_{j=1}^{d}\alpha_{ji}y_j\right)+c_i\right\},
	\end{equation}
	where $\psi$ is an activation function and $\theta=\left(\beta_1,\cdots,\beta_n,\alpha_{11},\cdots,\alpha_{dn},c_1,\cdots,c_n\right)\in\mathbb{R}^{2n+n(1+d)}$ is a vector of the neural network parameters. Assume the following:
	\begin{itemize}
		\item $\psi$ is in $C^2(\mathbb{R}^d)$, bounded and non-constant.
		\item $[0,T]\times D$ is compact.
		\item $\text{supp}\,\nu_1\subset D_T$ and $\text{supp}\,\nu_2\subset D$.
		\item The above PDE \eqref{eqn:para_pde} has a unique solution, where this solution belongs to both $C(\bar{D}_T)$ and $C^{1+\frac{\eta}{2},2+\eta}(D_T)$ for $0\leq\eta\leq 1$, and
		\begin{equation}
			\sup_{D_T}\left(\left|\nabla_{y} u(t,y)\right|+\left|\nabla_{y}^2 u(t,y)\right|\right)<\infty.
		\end{equation}
		\item $\hat{\gamma}(t,y,u,p)$ and $\frac{\partial\alpha_i(t,y,u,p)}{\partial p_j}$ for $1\leq i,j\leq d$ are locally Lipschitz continuous, where Lipschitz constant has a polynomial growth in $u$ and $p$.
		\item $\dfrac{\partial\alpha_i(t,y,u,p)}{\partial u_{y_j}}$ is bounded, for $1\leq i,j\leq d$.
	\end{itemize}
	Then there is a constant
	\begin{equation}
		K=K\left(\sup\limits_{D_T}\left|u\right|,\,\sup\limits_{D_T}\left|\nabla_y u\right|,\,\sup\limits_{D_T}|\nabla_y^{2}u|\right)>0,
	\end{equation}
	such that for arbitrary positive $\epsilon>0$, there is a DNN function $f$ in $\mathfrak{C}(\psi)=\bigcup\limits_{n=1}^\infty \mathfrak{C}^n(\psi)$ satisfying $J(f)\leq K\epsilon$.
\end{thm}
\medskip
\begin{proof}
	By Theorem 3 in \cite{hornik1991approximation}, for every $\epsilon>0$ and $u\in C^{1,2}([0,T]\times\mathbb{R}^d)$, there is a DNN function $f=f(t,y;\theta)$ in $\mathfrak{C}(\psi)$ such that
	\begin{equation} \label{2-dense}
		\sup_{D_T} |\partial_t u-\partial_t f|\,+\sup_{\bar{D}_T,0\leq j\leq2}|\partial_y^{(j)} u-\partial_y^{(j)} f|<\epsilon.
	\end{equation}
	Also we may assume for $C>0$, nonnegative constants $c_1,c_2,c_3$ and $c_4$,
	\begin{equation}
		|\hat{\gamma}(t,y,u,p)-\hat{\gamma}(t,y,v,q)|\leq C\left(|u|^\frac{c_1}{2}+|v|^\frac{c_2}{2}+|p|^\frac{c_3}{2}+|q|^\frac{c_4}{2}+1\right)(|u-v|+|p-q|),
	\end{equation}
	by the local Lipschitz continuity of $\hat{\gamma}(t,y,u,p)$ in $u$ and $p$. We abbreviate $u(t,y)$ and $f(t,y;\theta)$ for convenience. From the H\"{o}lder inequality with exponents $r_1$ and $r_2$,
	\begin{flushleft}
		$\displaystyle\int_{D_T} |\hat{\gamma}(t,y,f,\nabla_y f)-\hat{\gamma}(t,y,u,\nabla_y u)|^2 \, d\nu_1$ \\
		
		$\leq C\displaystyle\int_{D_T} (|f|^{c_1}+|u|^{c_2}+|\nabla_y f|^{c_3}+|\nabla_y u|^{c_4}+1)(|f-u|^2+|\nabla_y f-\nabla_y u|^2)\,d\nu_1$ \\
		
		$\leq C\left(\displaystyle\int_{D_T} (|f|^{c_1}+|u|^{c_2}+|\nabla_y f|^{c_3}+|\nabla_y u|^{c_4}+1)^{r_1}\,d\nu_1 \right)^\frac{1}{r_1}$ \\
		$\quad\times\left(\displaystyle\int_{D_T}(|f-u|^2+|\nabla_y f-\nabla_y u|^2)^{r_2}\,d\nu_1\right)^\frac{1}{r_2}$ \\
		
		$\leq C\left(\displaystyle\int_{D_T} (|f-u|^{c_1}+|\nabla_y f-\nabla_y u|^{c_3}+|u|^{c_1\vee c_2}+|\nabla_y u|^{c_3\vee c_4}+1)^{r_1}\,d\nu_1 \right)^\frac{1}{r_1}$ \\
		$\quad\times\left(\displaystyle\int_{D_T}(|f-u|^2+|\nabla_y f-\nabla_y u|^2)^{r_2}\,d\nu_1\right)^\frac{1}{r_2}$ \\
		
		$\leq C \left(\epsilon^{c_1}+\epsilon^{c_3}+\sup\limits_{D_T}|u|^{c_1\vee c_2}+\sup\limits_{D_T}|\nabla_y u|^{c_3\vee c_4}\right)\epsilon^2$.
	\end{flushleft}
	Each constant $C$ from the above inequalities may differ from each other. The last inequality holds because of \eqref{2-dense}.
	
	Also we may assume
	\begin{equation}
		\left|\dfrac{\partial\alpha_i(t,y,u,p)}{\partial p_j}-\dfrac{\partial\alpha_i(t,y,v,q)}{\partial q_j}\right|\leq C\left(|u|^\frac{c_1}{2}+|v|^\frac{c_2}{2}+|p|^\frac{c_3}{2}+|q|^\frac{c_4}{2}+1\right)(|u-v|+|p-q|),
	\end{equation}
	by the local Lipschitz continuity of $\frac{\partial\alpha_i(t,y,u,p)}{\partial p_j}$ in $u$ and $p$. For convenience, we denote
	\begin{equation}
		\xi(t,y,h,\nabla h,\nabla^2 h)=\sum_{i,j=1}^{d} \dfrac{\partial\alpha_i(t,y,h,\nabla h)}{\partial h_{y_j}} \partial_{{y_i}{y_j}}h(t,y).
	\end{equation}
	In spirit to the above procedure we used the H\"{o}lder inequality with exponents $p$ and $q$:
	\begin{flushleft}
	$\displaystyle\int_{D_T} |\xi(t,y,u,\nabla_y u,\nabla_y^2 u)-\xi(t,y,f,\nabla_y f,\nabla_y^2 f)|^2 \, d\nu_1$ \\
	
	$\leq\displaystyle\int_{D_T} \left|\sum_{i,j=1}^{d}\left(\dfrac{\partial\alpha_i(t,y,f,\nabla f)}{\partial f_{y_j}}-\dfrac{\partial\alpha_i(t,y,u,\nabla u)}{\partial u_{y_j}}\right)\partial_{{y_i}{y_j}}u(t,y)\right|^2 \, d\nu_1$ \\
	$\quad+ \displaystyle\int_{D_T} \left|\sum_{i,j=1}^{d}\dfrac{\partial\alpha_i(t,y,f,\nabla f)}{\partial f_{y_j}}(\partial_{{y_i}{y_j}}f(t,y;\theta)-\partial_{{y_i}{y_j}}u(t,y))\right|^2 \, d\nu_1$ \\
	
	$\leq C\displaystyle\sum_{i,j=1}^{d}\left(\displaystyle\int_{D_T}|\partial_{{y_i}{y_j}}u(t,y)|^{2p}\,d\nu_1\right)^\frac{1}{p} \left(\displaystyle\int_{D_T} \left|\dfrac{\partial\alpha_i(t,y,f,\nabla f)}{\partial f_{y_j}}-\dfrac{\partial\alpha_i(t,y,u,\nabla u)}{\partial u_{y_j}}\right|^{2q}\,d\nu_1\right)^\frac{1}{q}$ \\
	$\quad+\,C\displaystyle\sum_{i,j=1}^{d}\left(\displaystyle\int_{D_T}\left|\dfrac{\partial\alpha_i(t,y,f,\nabla f)}{\partial f_{y_j}}\right|^{2p}\,d\nu_1\right)^\frac{1}{p} \left(\displaystyle\int_{D_T}|\partial_{{y_i}{y_j}}f(t,y;\theta)-\partial_{{y_i}{y_j}}u(t,y)|^{2q}\,d\nu_1\right)^\frac{1}{q}$ \\
	
	$\leq C\displaystyle\sum_{i,j=1}^{d}\left(\displaystyle\int_{D_T} |\partial_{{y_i}{y_j}}u(t,y)|^{2p}\,d\nu_1\right)^\frac{1}{p} \left(\displaystyle\int_{D_T} (|f-u|^2+|\nabla_y f-\nabla_y u|^2)^{qr_2}\,d\nu_1 \right)^\frac{1}{qr_2}$ \\
	$\quad\times\left(\displaystyle\int_{D_T} (|f-u|^{c_1}+|\nabla_y f-\nabla_y u|^{c_3}+|u|^{c_1\vee c_2}+|\nabla_y u|^{c_3\vee c_4}+1)^{qr_1}\,d\nu_1 \right)^\frac{1}{qr_1}$ \\
	$\quad+\,C\displaystyle\sum_{i,j=1}^{d}\left(\displaystyle\int_{D_T} \left|\dfrac{\partial\alpha_i(t,y,f,\nabla f)}{\partial f_{y_j}}\right|^{2p}\,d\nu_1\right)^\frac{1}{p}\left(\displaystyle\int_{D_T}|\partial_{{y_i}{y_j}}f(t,y;\theta)-\partial_{{y_i}{y_j}}u(t,y)|^{2q}\,d\nu_1\right)^\frac{1}{q}$ \\
	
	$\leq C\epsilon^2$. 
\end{flushleft}
	To sum up, we finally obtain the following inequality:
	\begin{equation}
		\begin{aligned}
			J(f)&=\lVert \mathcal{G}[f] \rVert ^2 _{D_T,\nu_1} + \lVert f(T,y;\theta)-u_T (y) \rVert ^2 _{D,\nu_2} \\
			&=\lVert \mathcal{G}[f]-\mathcal{G}[g] \rVert ^2 _{D_T,\nu_1} + \lVert f(T,y;\theta)-u_T (y) \rVert ^2 _{D,\nu_2} \\
			&\leq\displaystyle\int_{D_T}\left(|\partial_t u-\partial_t f|^2+|\xi(t,y,u,\nabla u,\nabla^2 u)-\xi(t,y,f,\nabla f,\nabla^2 f)|^2\right)\,d\nu_1\\
			&\quad+ \displaystyle\int_{D_T}|\hat{\gamma}(t,y,f,\nabla_y f)-\hat{\gamma}(t,y,u,\nabla_y u)|^2\,d\nu_1 + \displaystyle\int_D |f(T,y;\theta)-u_T(y)|^2\,d\nu_2 \\
			&\leq K\epsilon^2
		\end{aligned}
	\end{equation}
	for some constant $K>0$.
\end{proof}

\subsection{Convergence of the DNN function to the solution of PDEs}
As we done in section A.1, consider the quasilinear parabolic PDE \eqref{eqn:para_pde} and the following loss functional
\begin{equation}
	J(f)=\lVert \mathcal{G}[f] \rVert ^2 _{D_T,\nu_1} + \lVert f(T,y;\theta)-u_T (y) \rVert ^2 _{D,\nu_2}.
\end{equation}
By Theorem $\ref{thm_no_1}$, there is a neural network $f^n$ such that $J(f^n)$ tends to $0$. Each $f^n$ satisfies the following:
\begin{equation}   
	\begin{aligned}
		\mathcal{G}[f^n](t,y)&=h^n(t,y), &(t,y) \in D_T, \\
		f^n(T,y)&=u_T^n(y), &y \in D, \label{eqn:para_pde_fn} \\
	\end{aligned}
\end{equation}
and
\begin{equation}
	\lVert h^n \rVert ^2 _{D_T,\nu_1} + \lVert u_T^n-u_T \rVert ^2 _{D,\nu_2}\rightarrow0\text{ as }n\rightarrow\infty.
\end{equation}
\begin{thm} \label{thm_no_2}
	Assume the following:
	\begin{itemize}
		\item $\lVert \alpha(t,y,u,p) \rVert \leq \mu(\lVert p\rVert+\kappa(t,y))$ for all $(t,y)\in D_T$, with $\mu>0$ and $\kappa\in L^2(D_T)$ being positive.
		\item $\alpha$ is continuously differentiable in $(y,u,p)$.
		\item Both $\alpha$ and $\gamma$ are Lipschitz continuous, uniformly on the following form of compact sets:
		\begin{equation}
			\left\{(t,y,u,p):t\in[0,T],\,y\in\bar{D},\,0\leq|u|\leq C,\,0\leq\lVert p \rVert\leq C\right\}.
		\end{equation}
		\item $\left\langle p, \,\alpha(t,y,u,p) \right\rangle \geq \nu \lVert p \rVert^2$ for some $\nu>0$.
		\item $\left\langle p_1-p_2,\,\alpha(t,y,u,p_1)-\alpha(t,y,u,p_2)\right\rangle>0$ for some $\nu>0$, for every $p_1,p_2\in\mathbb{R}^d$ with $p_1\neq p_2$.
		\item $|\gamma(t,y,u,p)|\leq\lVert p\rVert\lambda(t,y)$ for all $(t,y)\in D_T$, with $\lambda\in L^{d+2}(D_T)$ being positive.
		\item $u_T(y)\in C^{0,2+\xi}(\bar{D})$ for some $\xi>0$. Note that
		\begin{equation}
			\lVert u(y) \rVert_{C^{0,\beta}(\bar{D})} = \sup_{y\in\bar{D}}|u(y)|^{[\beta]}+\sup_{y_1,y_2\in\bar{D},y_1\neq y_2}\dfrac{|u(y_1)-u(y_2)|}{|y_1-y_2|^{\beta-[\beta]}}.
		\end{equation}
		\item  $u_T$ and $u_T'$ are bounded in $\bar{D}$. 
		\item $D\subset\mathbb{R}^d$ is bounded and open with boundary $\partial D\in C^2$.
		\item $(f^n)_{n\in\mathbb{N}}\in C^{1,2}(\bar{D}_T)$ and $(f^n)_{n\in\mathbb{N}}\in L^2(D_T)$.
	\end{itemize}
	Then
	\begin{enumerate}
		\item the PDE \eqref{eqn:para_pde} has a unique bounded solution
		\begin{equation}
			u\in C^{0,\delta,\frac{\delta}{2}}(\bar{D}_T)\cap W_0^{(1,2),2}(D_T^{\star}) \cap L^2\left(0,T;W_0^{1,2}(D)\right),\quad \delta>0,
		\end{equation}
		for any interior subdomain $D_T^{\star}\subset D_T$.
		\item $f^n\rightarrow u$ strongly in $L^\rho(D_T)$ for every $\rho<2$.
	\end{enumerate}
\end{thm}
\medskip
Note that in case of the class of quasilinear parabolic PDEs with boundary conditions, we should also consider the limiting process in the weak formulation of PDEs and use the Vitali's theorem. For more detail, see Appendix A in \cite{sirignano2018dgm}. See also \cite{boccardo2009summability}, \cite{magliocca2018existence}, \cite{di2011existence} and \cite{debnath2011nonlinear}.
\begin{proof}
	Existence, regularity and uniqueness for \eqref{eqn:para_pde} follows from Theorem 2.1 in \cite{porzio1999existence} and Theorem 6.3 to 6.5 of chapter V.6 in \cite{ladyzhenskaia1968linear}. Boundedness holds by Theorem 2.1 in \cite{porzio1999existence}. See also chapter V.2 from \cite{ladyzhenskaia1968linear}.
	
	Let $f^n$ be the solution of \eqref{eqn:para_pde_fn}. By Lemma 4.1 of \cite{porzio1999existence}, $\left\{f^n\right\}_{n\in\mathbb{N}}$ is uniformly bounded in both $L^\infty(0,T;L^2(D))$ and $L^2\left(0,T;W_0^{1,2}(D)\right)$. Then we can pick a subsequence from the sequence of neural networks $\left\{f^n\right\}_{n\in\mathbb{N}}$, where we denote also by $\left\{f^n\right\}_{n\in\mathbb{N}}$ for convenience, satisfying
	\begin{itemize}
		\item $f^n \xrightarrow{w*} u$ in $L^\infty(0,T;L^2(D))$,
		\item $f^n \rightarrow u$, weakly in $L^2\left(0,T;W_0^{1,2}(D)\right)$,
		\item $f^n(\cdot,t) \rightarrow v(\cdot,t)$, weakly in $L^2(D)$, for every fixed $t$ in $[0,T)$,
	\end{itemize}
	for some functions $u,v$.
	Since the norm of $f$ in a Banach space $L^2\left(0,T;W_0^{1,2}(D)\right)$ is defined as
	\begin{equation}
		\lVert f \rVert_{L^2\left(0,T;W_0^{1,2}(D)\right)} = \left(\displaystyle\int_{0}^{T}\lVert f\rVert_{W_0^{1,2}(D)}^2\,dt\right)^\frac{1}{2},
	\end{equation}
	where
	\begin{equation}
		\lVert f \rVert_{W_0^{1,2}(D)}^2 = \sum_{|\alpha|\leq2}\lVert D^{\alpha}f \rVert_{L^2(D)}^2 = \lVert f \rVert_{L^2(D)}^2 + \lVert Df \rVert_{L^2(D)}^2 + \lVert D^2f \rVert_{L^2(D)}^2,
	\end{equation}
	$\left\{\nabla_y f^n\right\}_{n\in\mathbb{N}}$ is uniformly bounded in $L^2(0,T;W_0^{1,2}(D))$.
	
	Let $q=1+\dfrac{d}{d+4}\in(1,2)$. By the H\"{o}lder inequality with exponents $r_1,r_2>1$,
	\begin{equation} 
		\begin{aligned}
			\displaystyle\int_{D_T}\left|\gamma(t,y,f^n,\nabla_y f^n)\right|^q \, dtdy &\leq \displaystyle\int_{D_T} |\lambda(t,y)|^q |\nabla_y f^n(t,y)|^q \, dtdy \\
			&\leq \left(\displaystyle\int_{D_T}|\lambda(t,y)|^{r_1 q}\,dtdy\right)^\frac{1}{r_1}\left(\displaystyle\int_{D_T}|\nabla_y f^n(t,y)|^{r_2 q}\,dtdy\right)^\frac{1}{r_2}. \label{ineq_gamma}
		\end{aligned}
	\end{equation}
	Choose $r_2=\dfrac{2}{q}$. Then we get $r_1=\dfrac{2}{2-q}$ and hence $r_1 q=d+2$. Since $\lambda\in L^{d+2}(D_T)$ and $\left\{\nabla_y f^n\right\}_{n\in\mathbb{N}}$ is uniformly bounded,
	\begin{equation}
		\displaystyle\int_{D_T}\left|\gamma(t,y,f^n,\nabla_y f^n)\right|^q\,dtdx \leq C,
	\end{equation}
    for some $C>0$.
	
	The growth assumption on $\alpha$ and the above argument imply that $\left\{\partial_t f^n\right\}_{n\in\mathbb{N}}$ is uniformly bounded in $L^{1+\frac{d}{d+4}}(D_T)$ and $L^2\left(0,T;W^{-1,2}(D)\right)$. Let $\delta_1,\delta_2$ be the conjugate exponents satisfying $\delta_2>\max\left\{2,d\right\}$. By the Gagliardo--Nirenberg--Sobolev inequality and the Rellich--Kondrachov compactness theorem(for further details, see chapter 5 in \cite{evans2002partial}), the following embeddings hold:
	\begin{equation}
		W^{-1,2}(D) \subset W^{-1,\delta_1}(D), \quad L^q(D) \subset W^{-1,\delta_1}(D)\text{, and  }L^2(D) \subset W^{-1,\delta_1}(D),
	\end{equation}
	and hence $\left\{\partial_t f^n\right\}_{n\in\mathbb{N}}$ is uniformly bounded in $L^1(0,T;W^{-1,\delta_1}(D))$.
	
	By Corollary 4 in \cite{simon1986compact} and the following embedding
	\begin{equation}
		W_0^{1,2}(D) \subset\subset L^2(D) \subset W^{-1,\delta_1}(D),
	\end{equation}
	$\left\{f^n\right\}_{n\in\mathbb{N}}$ is relatively compact in $L^2(D_T)$, in other words,
	\begin{equation}
		f^n \rightarrow u\text{ strongly in }L^2(D_T)\text{ as }n\rightarrow\infty.
	\end{equation}
	Thus
	\begin{equation}
		f^n \rightarrow u \text{ almost everywhere in }D_T\text{ up to subsequences.} \label{final_pf_1}
	\end{equation}	
	Note that from the Theorem 3.3 of \cite{boccardo1997nonlinear}, we get
	\begin{equation}
		\nabla f^n \rightarrow \nabla u\text{ almost everywhere in }D_T. \label{final_pf_2}
	\end{equation}
	Hence $f^n \rightarrow u$ strongly in $L^{\rho}\left(0,T;W_0^{1,\rho}(D)\right)$ and so in $L^{\rho}(D_T)$ for every $\rho<2$, by \eqref{final_pf_1} and \eqref{final_pf_2}.

\end{proof}

\end{appendices}

\noindent\textbf{Acknowledgement.}\\ 
Hyungbin Park was supported by Research Resettlement Fund for the new faculty of Seoul National University.
Hyungbin Park was also supported by the National Research Foundation of Korea (NRF) grants funded by the Ministry of Science and ICT  (No. 2018R1C1B5085491 and No. 2017R1A5A1015626) 
and the Ministry of Education   (No. 2019R1A6A1A10073437) through Basic Science Research Program.

\bibliographystyle{apa}

\bibliography{HJB_paper}
\nocite{al2018solving, bjork2009arbitrage, crisostomo2014analyisis, guasoni2015static, mehrdoust2020calibration, remani2013numerical, sirignano2018dgm}

\end{document}